\documentclass[a4paper,fleqn]{cas-sc}

\usepackage[authoryear]{natbib}
\usepackage{amssymb}
\usepackage{multirow}
\usepackage{subfigure}
\usepackage{algorithm}
\usepackage{algorithmic}
\usepackage{colortbl}
\usepackage{xcolor}
\usepackage{dcolumn}
\usepackage{caption}
\usepackage{tabularx}
\usepackage{booktabs}
\usepackage{soul}
\usepackage{amsmath,mathtools}

\def\tsc#1{\csdef{#1}{\textsc{\lowercase{#1}}\xspace}}
\tsc{WGM}
\tsc{QE}
\tsc{EP}
\tsc{PMS}
\tsc{BEC}
\tsc{DE}

\begin{document}
\newsavebox\CBox
\def\textBF#1{\sbox\CBox{#1}\resizebox{\wd\CBox}{\ht\CBox}{\textbf{#1}}}

\let\WriteBookmarks\relax
\def\floatpagepagefraction{1}
\def\textpagefraction{.001}

\shorttitle{Predicting Viral Rumors and Vulnerable Users for Infodemic Surveillance}

\shortauthors{Zhang et~al.}

\title [mode = title]{Predicting Viral Rumors and Vulnerable Users with Graph-based Neural Multi-task Learning for Infodemic Surveillance}                      



%

\author[1]{Xuan Zhang}[orcid=0009-0000-8760-4381]


\ead{xuanzhang.2020@phdcs.smu.edu.sg}
\cormark[1]

\credit{
Conceptualization, Methodology, Software, Validation, Formal analysis, Investigation, Data curation, Writing– original draft}
\affiliation[1]{organization={School of Computing and Information Systems, Singapore Management University},
    addressline={80 Stamford Rd}, 
    country={Singapore}}

\author[1]{Wei Gao}[orcid=0000-0003-2028-2407]
\ead{weigao@smu.edu.sg}

\credit{Conceptualization, Investigation, Methodology, Writing– review \& editing, Supervision, Funding acquisition}

\cortext[cor1]{Corresponding author}



\begin{abstract}
In the age of the infodemic, it is crucial to have tools for effectively monitoring the spread of rampant rumors that can quickly go viral, as well as identifying vulnerable users who may be more susceptible to spreading such misinformation. This proactive approach allows for timely preventive measures to be taken, mitigating the negative impact of false information on society.
We propose a novel approach to predict viral rumors and vulnerable users using a unified graph neural network model. We pre-train network-based user embeddings and leverage a cross-attention mechanism between users and posts, together with a community-enhanced vulnerability propagation (CVP) method to improve user and propagation graph representations. Furthermore, we employ two multi-task training strategies to mitigate negative transfer effects among tasks in different settings, enhancing the overall performance of our approach.
We also construct two datasets with ground-truth annotations on information virality and user vulnerability in rumor and non-rumor events, which are automatically derived from existing rumor detection datasets. 
Extensive evaluation results of our joint learning model confirm its superiority over strong baselines in all three tasks: rumor detection, virality prediction, and user vulnerability scoring. For instance, compared to the best baselines based on the Weibo dataset, our model makes 3.8\% and 3.0\% improvements on Accuracy and MacF1 for rumor detection, and reduces mean squared error (MSE) by 23.9\% and 16.5\% for virality prediction and user vulnerability scoring, respectively. Our findings suggest that our approach effectively captures the correlation between rumor virality and user vulnerability, leveraging this information to improve prediction performance and provide a valuable tool for infodemic surveillance.

\end{abstract}




\maketitle

\section{Introduction}
\label{intro}

Online rumors refer to unverified information circulating on the Internet, especially on social media. Rumors with misinformation that distort facts can cause false beliefs and unnecessary panic in public.
Those viral rumors, specifically, can result in incredibly harmful repercussions due to their extensive reachability, evidenced as various man-made tragedies sparked by the rumors about COVID-19 during the latest pandemic~(\citep{islam2021covid,ali2020covid}.

Automatic rumor detection based on machine learning methods is an active research topic in recent years~\citep{ma2016detecting,ma2017detect,bian2020rumor,ma-etal-2018-rumor,2020fang}.
However, the potential for online rumor surveillance is limited by the cost-effectiveness of monitoring all identified rumors and participants indiscriminately, while not all rumors and participants are equally important in surveillance.
For example, a widely spreading rumor can have a greater range of impact and tends to be more convincing to users since high virality can serve as a mental shortcut to make a judgment by evoking a greater perception of social norms~\citep{kim2018rumor,lee2017normative}; meanwhile, a rumor can affect individual recipients differently, depending on how gullible each recipient is~\citep{mercier2017gullible}, which may in turn influence how broadly and deeply the rumor diffuses.
By accurately estimating information \textbf{virality} as well as user \textbf{vulnerability} in the context of rumor detection, we can expect to track noteworthy rumors and users more cost-effectively on various social media platforms.

\begin{figure}[t!]
\centering
\includegraphics[width=0.8\textwidth]{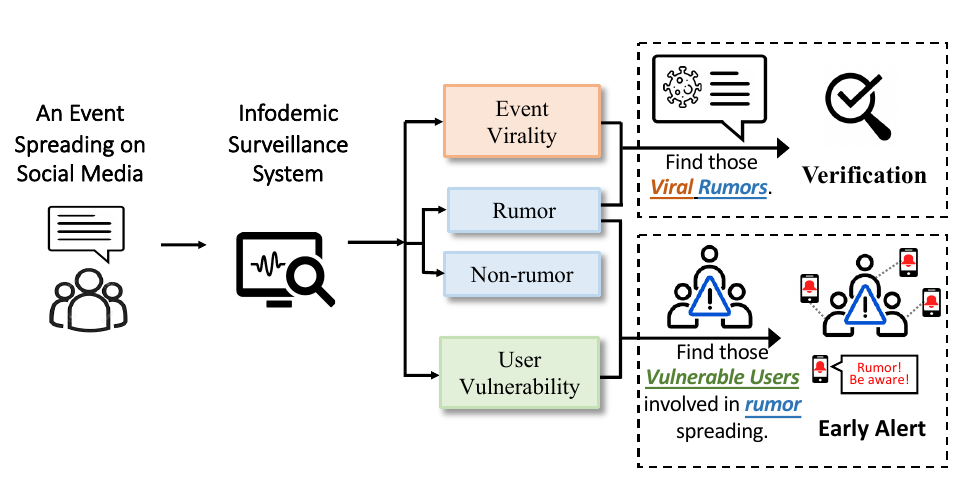} 
\caption{An application scenario of the infodemic surveillance system that can predict viral rumors and vulnerable users.}
\label{fig:application}
\end{figure}

However, such needs currently exceed the capacity of various existing rumor detection tasks as they generally overlook the assessment of the potential reach of rumors and the hazard of their spreaders. Imagine if there were a surveillance system that combined rumor detection, virality prediction, and vulnerability prediction as illustrated in Figure~\ref{fig:application}, with such a system, stakeholders would be able to send alert to those at-risk vulnerable users and request the potentially viral rumors to be professionally verified in a timely manner, resulting in more effective information surveillance and early intervention. To meet this goal, it is necessary to bridge a few major gaps.

Previous information diffusion research has investigated the virality prediction of general information. For example,~\citet{Cheng_2014} combined various temporal and structural features with machine learning models to estimate how large an information propagation graph of a source message can reach in the future.~\citet{li2016deepcas} used neural network models to learn feature representations that characterize virality for a similar purpose. However, such general virality models do not consider the distinction between rumors and normal information, and the influence of gullible users on diffusion, who are more likely to be activated by rumors.

Among misinformation related studies, the concept of user vulnerability (i.e., gullibility) is proposed to measure the propensity that motivates a user to participate in spreading rumor or fake news. For example,~\citet{10.1145/3292522.3326055} attempt to predict user vulnerability to fake news using a classifier that considers features of post content, user status (e.g., number of followers and friends), and network connectivity. However, these works do not further link user vulnerability to rumor's virality.
In social network analysis, the susceptibility of users to item adoption is measured along with topology intrinsic factors~\citep{hoang2012virality}, such as the ability of users to spread items and the potential of users to adopt items, which contribute to the wide dissemination of items in the network. Although this may help analyze the relationship between information virality and the strength of users to spread (or adopt) information from a topology perspective, rumor virality is arguably not only related to the network's topological structure but also to the content of the information being spread and user's cognitive state.

\begin{figure*}
\centering

\includegraphics[width=1\textwidth]{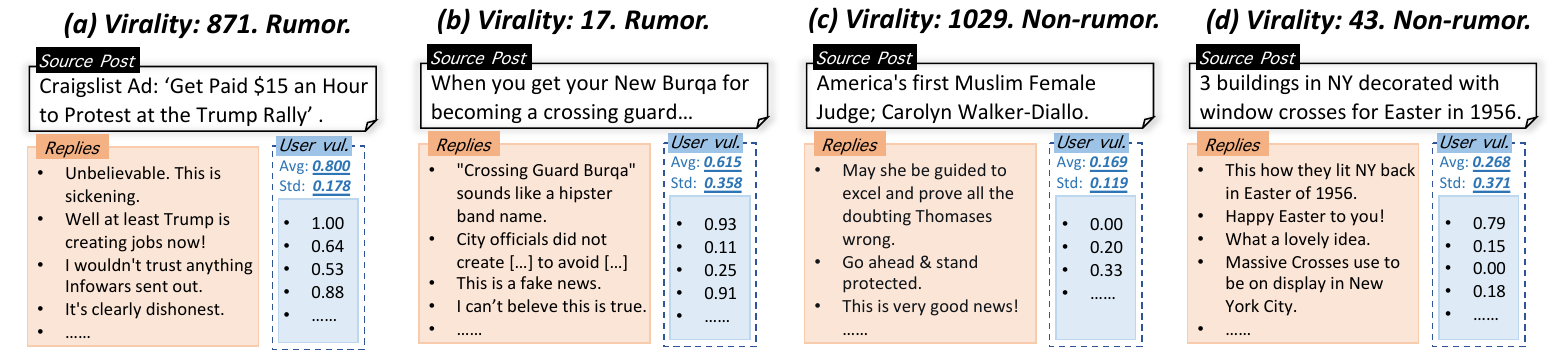} 

\caption{Example rumors and non-rumors of different virality and reposts to them by users of different vulnerabilities, taken from TWITTER dataset~\citep{ma2017detect}. Virality is defined as the number of users involved in the spread, and user vulnerability is defined as the fraction of rumor events over all events a user engaged in.}
\label{fig:intro-exapmle}
\end{figure*}

Based on public rumor detection datasets such as TWITTER\footnote{We combine the Twitter15 and Twitter16 datasets released by~\citet{ma2017detect} into one larger dataset, namely TWITTER.}~\citep{ma2017detect}, which typically consist of a set of propagation cascades of source posts labeled as rumor and non-rumor, 
we observe that the virality (i.e., the number of users participating in the spread) can be in some extent related to the interaction between the vulnerability of the involved users (i.e., user's propensity of engaging in rumor spreading)\footnote{The user's vulnerability is estimated as the proportion of rumors among all the information, in which the user is engaged.} and whether the information propagated is a rumor or a non-rumor.
Figure~\ref{fig:intro-exapmle} presents four illustrative examples from the TWITTER dataset. 
Overall, most users attracted by non-rumors are typically much less vulnerable than in rumors.
Notably, rumors tend to spread more virally when the involved users are overall more vulnerable as shown in (a) vs (b). Meanwhile, non-rumors tend to be more viral when the participating users are generally less vulnerable, as illustrated by the distinction between (c) and (d). 
According to the above observations, it seems that some implicit links exist among the user vulnerability, the virality and the rumor/non-rumor nature of concerned information. If this held true, it would be likely for us to harness their relationship to facilitate three prediction tasks concerning rumor detection, information virality prediction, and user vulnerability estimation altogether. One might typically assume that virality is irrelevant to whether the concerned information is a rumor or not because both rumors and non-rumors could be viral (or not viral). However, we argue that given user vulnerability as a bridge, the relationship between the virality and the rumor/non-rumor nature of information could be established implicitly and validated with the performance of the relevant prediction tasks. 

To verify our presumption, we propose a unified multi-task learning framework based on Graph Neural Networks (GNN) and hierarchical graph pooling methods for joint learning of rumor detection, virality prediction, and user vulnerability scoring. The framework aims to improve the performance of the three seemingly independent tasks simultaneously to aid information surveillance and early intervention.
In particular, we first transform information spreading networks into user interaction networks, which are used to train user embeddings using a network-based approach. The user embeddings are further enhanced by incorporating post content using a user-post cross-attention mechanism. Then we use DiffPool~\citep{ying2019hierarchical} to discover latent communities of users that exhibit similar behavior in spreading rumors, and use a Community-enhanced Vulnerability Propagation (CVP) method with reference to the latent communities to refine user embeddings. 
To mitigate negative transfer among the tasks, we adopt two radically different multi-task training methods, Gradnorm~\citep{chen2018gradnorm} and meta-learning~\citep{buffelli2020meta} over the three tasks.
Our main contributions can be summarized as follows:
\begin{itemize}
\item To our best knowledge, this is the first study on rumor detection, virality prediction, and user vulnerability scoring in a unified framework for infodemic surveillance.
\item We use inductive GNNs and hierarchical graph pooling to learn the three tasks jointly. In particular, we pre-train general user embeddings and propose a CVP method to further refine user embeddings
\item We train the framework under two  different multi-task settings, namely concurrent training with GradNorm and meta-learning, for dealing with task training conflicts, i.e., negative transfer.
\item We build two large datasets with virality and user vulnerability annotated based on existing rumor detection datasets, on which our method outperforms strong baselines for all three tasks\footnote{Data and code are available at \url{https://github.com/jadeCurl/Predicting-Viral-Rumors-and-Vulnerable-Users}.}.
\end{itemize}

\section{Related Work}
Our study in this paper intersects multiple research topics including rumor detection, information virality prediction and user vulnerability analysis under multi-task learning framework, which are reviewed separately in this section.

\subsection{Rumor Detection}

Several surveys~\citep{cao2018automatic,zubiaga2018detection,zannettou2019web,sharma2019combating,xu2021unified} have comprehensively reviewed the literature on rumor detection. Here we introduce some works most relevant to us.
Many studies on rumor detection mainly extract discriminative features from post content, user profiles, and propagation patterns, and learn supervised classifiers for classification~\citep{castillo2011information,zhao2015enquiring,ma2017detect}. 
Later, representation-learning methods become a dominant approach by exploiting end-to-end neural network models to learn latent representations for classification~\citep{ma-etal-2018-rumor,bian2020rumor,nguyen2020fang,liu2020fned,alkhodair2020detecting,dou2021user,bi2022microblog,sun2022ddgcn}. For example, ~\citet{bian2020rumor} use a Bi-directional Graph Convolutional Network (Bi-GCN) to capture the diffusion and dispersion properties from propagation trees. ~\citet{sun2022ddgcn} propose a Dual-Dynamic Graph Convolutional Networks (DDGCN) model to learn the dynamics of messages in propagation and the dynamics of background knowledge from Knowledge graphs simultaneously.
However, given the multitude of rumors online, it is not cost-effective nor necessary to track all the rumors detected. 
Differing from existing rumor detection methods which aim to distinguish rumors from non-rumors, we allow the systems to be able to distinguish the virality of rumors and non-rumors in terms of the number of \emph{unique} users that they may reach for more cost-effective surveillance. 

\subsection{Information Virality Prediction}

To the best of our knowledge, there is little work focusing on predicting the virality of rumors. However, general studies on the virality of information diffusion are abundant~\citep{articleyu,Zhao_2015,li2016deepcas,8554730,chen2021catch,zhang2021deepblue,tan2022efficient}.
Existing approaches are commonly based on feature engineering using original post, sharer, and community structure~\citep{10.1145/2487788.2488017,Cheng_2014, weng2014predicting}. 
Based on temporal and structural features, generative models were also designed~\citep{shen2014modeling,articleyu}. For example,~\citet{shen2014modeling} modeled the information diffusion as a reinforced Poisson process. Those generative approaches made strong assumptions over macroscopic distributions and stochastic processes, which limits their application in practice.
For better feature representation, deep neural networks were utilized with promising results for virality prediction~\citep{li2016deepcas,8554730,zhang2021deepblue,tan2022efficient}.

However, these general virality prediction works do not take into account the interaction between vulnerability of involved users and whether the information propagated is a rumor or a non-rumor. To the best of our knowledge, there is no existing research specifically studying the difference between the virality of rumors and that of non-rumor information. In the fake news domain, some works found that fake news spreads faster and wider than true news in general~\citep{vosoughi2018spread}. However, no work has been attempted to predict the extent of rumor's propagation directly, and it remains unclear how such extent differs between rumor and non-rumor. Our study does not discriminately predict the virality of rumor and non-rumor either. Instead, our prediction of viral rumors is achieved indirectly by cross-referencing the outputs of the rumor detection task and the virality prediction task, which not only share common features at the graph level but also have the user vulnerability task as the bridge. As a result, we can better predict the virality of information if the information type (i.e., rumor or non-rumor) and the vulnerability traits of participating users are known. This is useful from an information surveillance point of view because the joint prediction allows us to keep track on those widely circulated rumors and vulnerable users at same time.

\subsection{User Vulnerability Analysis}
Existing works have been focused on studying users' susceptibility to various information sources in social media and social networks~\citep{6642447,lee2015measuring,hoang2012virality,vuln-Al}.
There have been few previous analyses on the vulnerability of users to rumors and fake news~\citep{10.1145/3341161.3342920,pennycook2019lazy,bringula2021gullible}.  \citet{10.1145/3341161.3342920} proposed a community health assessment model based on the concept of believability derived from computational trust metrics to calculate the vulnerability of nodes and communities to fake news spread.  \citet{bringula2021gullible} found that technological, internal, and external factors may positively or negatively affect university students’ vulnerability to political misinformation.  In this paper, we use a unified neural model to learn features representing user vulnerability to rumors and further link user vulnerability to the virality of information, that can
help rumor detection and the estimation of extent of rumors' potential reach
to network users.

\subsection{Graph Neural Networks (GNNs)}
Recently, Graph Convolutional Networks (GCNs) based on matrix factorization have demonstrated state-of-the-art performance in various graph-related tasks~\citep{2018graph,sun2022ddgcn,warmsley2022survey,zhang2023detecting,he2022graph,zhai2023causality,wang2023jointly}. These transductive GCNs work well on fixed graphs, but are difficult to  apply to unseen nodes that do not appear at training time~\citep{defferrard2017convolutional}. To overcome this weakness, \citet{hamilton2018inductive} proposed an inductive GraphSAGE model which aggregates node features from neighbors. 
Different kinds of hierarchical pooling processes can be added on top of the GNN models to capture the deeper information from hierarchical graph structures, for strengthening the learned node embedding~\citep{zhang2019hierarchical}.  For example, gPool~\citep{gao2019learning} and SAGPoo~\citep{lee2019self} select top-K nodes to form induced subgraphs, which may lose useful graph structure and node information. EdgePool~\citep{diehl2019edge} 
reduces edges in the graph to obtain pooled subgraphs by fixing the number of nodes in the pooled graph to half of the original. 
The above methods can learn a pooled smaller graph by actively dropping some node and edge information. 
In contrast, DiffPool~\citep{ying2019hierarchical}  map the nodes into a set of communities via soft assignments. In this paper, we use GraphSAGE and DiffPool to build our multi-task learning framework.

\section{Problem Definition}
The spread of rumors, e.g., ``Bat soup caused Wuhan virus'', on social media platforms can cause significant harm. Disseminating such false information may arouse misperception, hatred, and even ethnic conflict. We aim to facilitate the monitoring and early warning of the rampant spread of rumors by utilizing initial propagation information. Specifically, we focus on a series of prediction objectives: (1) differentiat rumors and non-rumors given an early (thus partial) observation of information propagation on social media platforms, (2) predict the scale of propagation of a concerned message given the observation of propagation at initial stages, (3) identify those vulnerable users who appear in the observed propagation and are more likely to disseminate such rumors, and assign them a vulnerability score, and (4) strengthen the performance of joint prediction on rumor, virality, and user vulnerability through improved representations of task-relevant features. 

To address these issues, we define our research problems as follows. 
Given a rumor dataset, each instance is given as a propagation network corresponding to a specific social event (or claim), which consists of a source post and its cascading messages spreading the event via reposting (e.g., replying and retweeting) behaviors~\citep{bian2020rumor,song2021adversary}. We define an entire post propagation network as graph $\mathrm{G}$, in which the timestamp of the last post is $T$. Additionally, given the predictive nature of our task, which requires information from the early stages of post propagation, we introduce an observed (i.e., partial) propagation network $\mathcal{G}\subset \mathrm{G}$, where $\mathcal{G}=\{\mathbf{V}, \mathbf{A}\}$ comprises a set of nodes and edges:
\begin{itemize}
\item $\mathbf{V}=\{ v_1,v_2,\ldots,v_{|\mathbf{V}|}\}$ is the node set corresponding to the posts engaging in the propagation. Each $v_i = (s_i,t_i,u_j)$ indicates that it is the $j$-th user $u_j\in \mathbf{U}$ ($1 \le j \le |\mathbf{U}|$) who creates the $i$-th post content $s_i$ ($1 \le i \le |\mathbf{V}|$) at timestamp $t_i$ ($t_i\in[0,t]$, where $t$ is the size of observation time window for $\mathcal{G}$ and $t<T$), and $\mathbf{U}=\{u_1,u_2,\ldots,u_{|\mathbf{U}|}\}$ is the set of unique users that appear in $\mathcal{G}$.
\item $\mathbf{A}^{|\mathbf{V}|\times|\mathbf{V}|}$ is the adjacency matrix indicating the reposting relationships among the posts. Each entry in the matrix is a binary value (0 or 1) that indicates whether there is a reposting relationship between two corresponding posts. Without the loss of generality, we omit the direction of message propagation, i.e., $\mathbf{A}$ is symmetric.
\end{itemize}

We then formulate three prediction tasks as below, corresponding to our first three research objectives:
\begin{enumerate}
    \item \emph{Rumor Detection:} Given an observed propagation network $\mathcal{G}\subset \mathrm{G}$, we formulate this task as a two-way graph classification problem of predicting whether the final propagation network $\mathrm{G}$ will be a rumor or not, namely the task $\mathcal{T}_1: \mathcal{G} \rightarrow y(\mathrm{G})\in\{\text{rumor},\text{non-rumor}\}$, where $y(\mathrm{G})$ is the ground-truth class of $\mathrm{G}$.
    \item \emph{Virality Prediction:} Given an observed propagation network $\mathcal{G}\subset\mathrm{G}$, we define this task as a regression problem of forecasting the total number of unique users participating in spreading the entire event $\mathrm{G}$, denoted as $\mathcal{T}_2: \mathcal{G} \rightarrow  \log_2|\mathbf{U}_\mathrm{G}|$, where $\mathbf{U}_\mathrm{G}$ represents the set of unique users that appear in $\mathrm{G}$. Following previous work~\citep{tsur2012s,kupavskii2012prediction,li2016deepcas}, the model will fit the logarithm of ground-truth $|\mathbf{U}_\mathrm{G}|$ by squashing the absolute size of network. 
    \item \emph{User Vulnerability Scoring:} Given an observed propagation network $\mathcal{G} \subset \mathrm{G}$, we try to infer the vulnerability of each unique user $u_j\in \mathcal{G}$ as how much it is susceptible to rumor spreading, which is denoted as the task $\mathcal{T}_3: u_j \rightarrow [0,1]$. The ground-truth user vulnerability is defined as the fraction of rumors over all the information in which the user is engaged.
\end{enumerate}
Furthermore, in order to address the fourth research objective, we propose a unified GNN model and adopt multi-task learning frameworks to jointly train the model for the three tasks above. To mitigate training conflicts among these tasks, we leverage the underlying correlations that might suggest their mutual predictability, which allows us to effectively integrate and optimize the prediction of rumor, virality, and user vulnerability within a single model.

\section{Methodology}

\begin{figure*}[htb]
\centering
\includegraphics[width=1\textwidth]{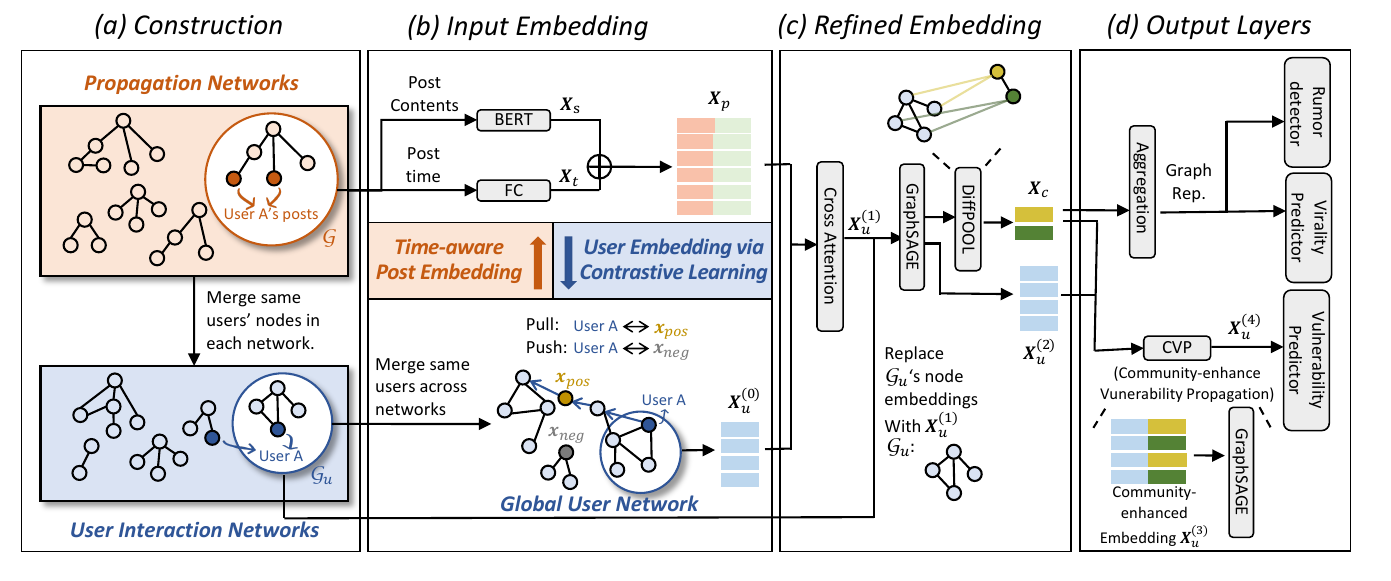} 
\caption{Overview of the proposed multi-task model. (a) Construction. The user interaction network $\mathcal{G}_u$ is constructed based on its corresponding post propagation network $\mathcal{G}$. (b) Input Embedding. We generate time-aware post embedding $\mathbf{X}_p$, and general user embedding $\mathbf{X}_u^{(0)}$. (c) Refined Embedding. We obtain latent user community information $\mathbf{X}_c$ via Diffpool. (d) Output Layers. The final graph representation for $\mathcal{G}_u$ and user representations $\mathbf{X}_u^{(4)}$ updated via CVP are fed to the corresponding classifiers for our three tasks. \emph{Note that the three tasks become separated only at (d) Output Layer, while the other layers (a) Construction, (b) Input Embedding, and (c) Refined Embedding are all needed for the three tasks as parts of their inputs and representation learning process}}.
\label{framework1}
\end{figure*}

As shown in Figure~\ref{framework1}, our whole framework consists of four components: (i) user interaction graph construction; (ii) input embedding; (iii) refined embedding; and (iv) output layers.

\subsection{User Interaction Graph Construction}
\begin{figure}[tbh]
\centering

\includegraphics[width=0.6\textwidth]{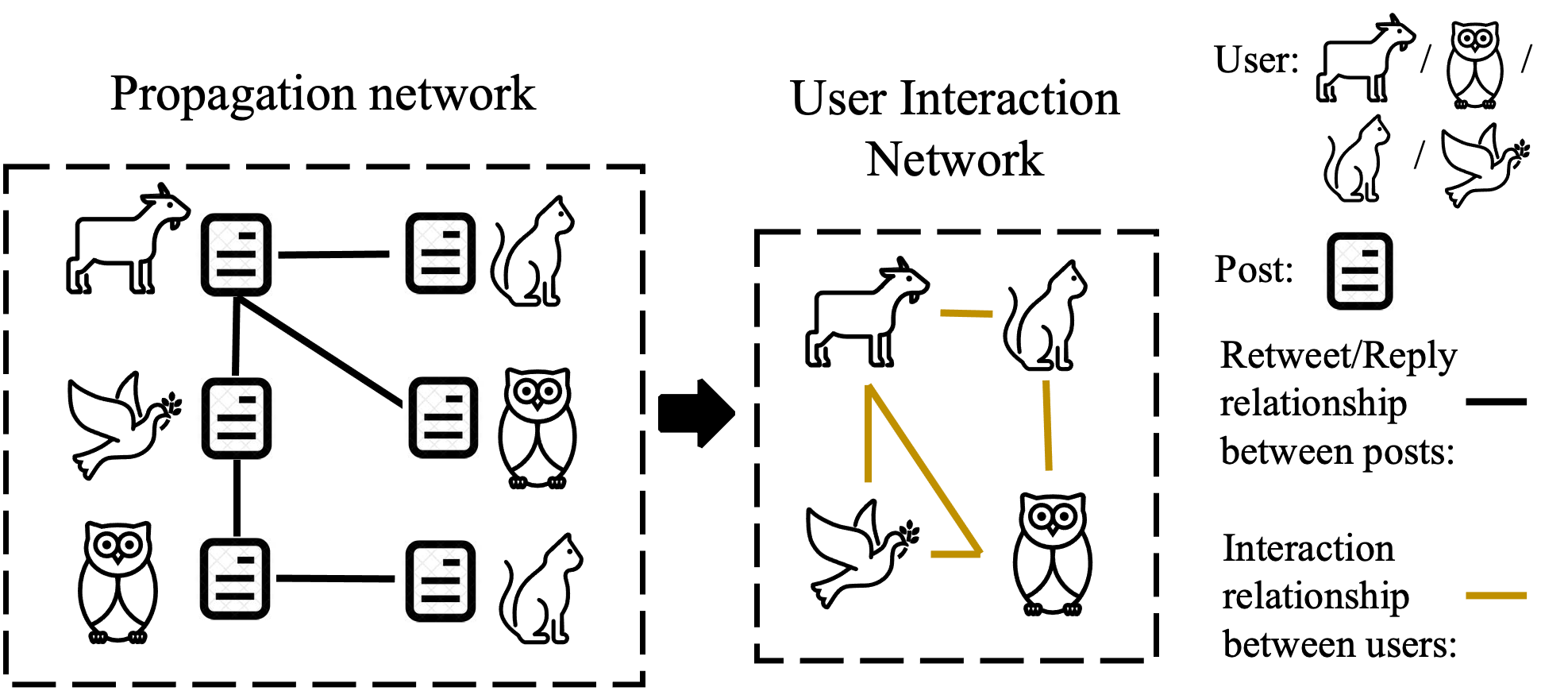}

\caption{An illustration of user interaction network construction. For any pair of unique users, we create an edge between them in the user interaction network as long as there is a reposting behavior between their posts in any post propagation network.} 
\label{fig:user}
\end{figure}

To understand the patterns of how users spread rumors and non-rumors, predict the likelihood of a source post going viral, and identify users who are more vulnerable to spreading misinformation, we first need to find the users involved in the propagation and the relationships between them.
Similar to existing work~\citep{ratkiewicz2011truthy,li2016deepcas}, we construct user interaction graphs $\mathcal{G}_u = \{\mathbf{U}, \mathbf{A}_u\}$ where the nodes are individual users based on propagation network $\mathcal{G}$, as shown in Figure~\ref{fig:user}. Basically, for any pair of unique users, we create an edge between them in the user interaction network as long as there is a reposting behavior between their posts in any post propagation network.
Formally, we denote $\mathbf{A}_u \in \mathbb{R}^{|\mathbf{U}|\times|\mathbf{U}|}$ is the adjacency matrix indicating relationships among the users. Compared to a follower/followee network, the user interaction network structure reflects the truly active interactions between users.

\subsection{Input Embedding}
The input embedding layer is used to capture the primitive characteristics of each user. To achieve this, the layer is divided into two separate modules: the time-aware post embedding and the contrastive learning-based user embedding as shown in Figure~\ref{framework1}(b). 
The main idea is to use the post information and users' global interactions to improve the user embeddings, enabling a more accurate representation of the users and their behavior.

\subsubsection{Time-aware Post Embedding}

Post information can reveal users' inner characteristics~\citep{rissola2019personality} and opinions~\citep{zhu2015pareto}, which is important in building user representation and can further help tasks like rumor detection~\citep{bian2020rumor} and virality prediction~\citep{chen2019npp}.
To obtain the representation of each post's content, we use a pre-trained BERT-based encoder~\citep{devlin2018bert}.
Specifically, we first flatten the propagation network  $\mathcal{G}$ in chronological order, which consist of a sequence of posts. This allows each post representation to jointly attend to nodes in different positions for better capturing semantics. Secondly, we insert two special tokens, i.e., [CLS] and [SEP], to the beginning and the end of post content $s_i \in v_i$, where the [CLS] token is intended to represent the semantic meaning of the post following it. 
The feature of token [CLS] from the last layer is taken to represent $s_i$, which is denoted as $\mathbf{x}_{s_i} \in \mathbb{R}^d$.
To make the model more efficient, we froze BERT parameters when training our own model.

Apart from the content feature, the time of each post is expected helpful in signaling the information type~\citep{nguyen2020fang} and future virality~\citep{8846015}. Thus, we make post embedding time-dependent by using a fully connected (FC) layer to convert $t_i \in v_i$ into a vector $\mathbf{x}_{t_i}$ and create the time-aware post embedding as $\mathbf{x}_{p_i}=[\mathbf{x}_{s_i},\mathbf{x}_{t_i}]$, where $[,]$ is the concatenation operator. Let matrix $\mathbf{X}_p\in \mathbb{R}^{\mathbf{|V|}\times2d}$ denote embedding of all the posts in propagation network $\mathcal{G}$.

\subsubsection{Pre-train User Embedding with Contrastive Learning}

The information diffusion process naturally reflects the rich proximity relationships between users~\citep{zhang2018cosine}. Mining the deep structure and patterns hidden in such relational networks can help learn the representation of users~\citep{pan2019social}. In this paper, we first construct a global user network across specific events based on the information diffusion process, and then study the general characteristics of users by their locations and connections in the global user network.
              
Specifically, we first create a global user graph by merging the same users across different user interaction graphs into one node. 
We do not assume any user-level annotation is available for it.
Then we use self-supervised contrastive learning to learn user embeddings. Given a user $u_j$, we simulate a random walk of fixed length in the network to get the positive samples from the nodes encountered along the path, and randomly pick negative samples. Due to the large size of the user network, we restrict both the number of positive and negative samples as one. Let $\mathbf{x}_{u_j}^{(0)} \in \mathbb{R}^d$ denote $u_j$'s embedding, the objective of contrastive learning is as follows:
\begin{equation}
    L = \mathbf{x}_{u_j}^{(0)}\mathbf{x}_{pos}-\mathbf{x}_{u_j}^{(0)}\mathbf{x}_{neg},
\end{equation}
where $\mathbf{x}_{pos}$ and $\mathbf{x}_{neg}$ are embeddings of $u_j$'s positive and negative samples in the random walk, respectively.

\subsection{Refined Embedding}
For each event, we get both the user embeddings $\mathbf{X}_{u}^{(0)} \in \mathbb{R}^{|\mathbf{U}|\times d}$ and time-aware post embeddings $\mathbf{X}_{p} \in \mathbb{R}^{|\mathbf{V}|\times 2d}$ from the input embedding layer. 
Then we use a user-post cross attention~\citep{vaswani2017attention} mechanism by treating $\mathbf{X}_{u}^{(0)}$ as query, and $\mathbf{X}_{p}$ as key and value. This basically uses user-specific features in $\mathbf{X}_{u}^{(0)}$ to guide our model in extracting post features in $\mathbf{X}_{p}$ relevant to the corresponding users, and thus refine the user embeddings with the post embeddings, which is formulated as:
\begin{equation}
\begin{gathered}
Q=\mathbf{X}_{u}^{(0)} W_q, ~~K=\mathbf{X}_{p} W_k, ~~V=\mathbf{X}_{p} W_v, \\
\mathbf{X}_{u}^{(1)}=\operatorname{softmax}\left(\frac{Q K^T}{\sqrt{d}} V\right),
\end{gathered}
\end{equation}
where $\mathbf{X}_{u}^{(1)}$ represents the updated user embeddings with post information, and $\mathbf{W}_q$, $\mathbf{W}_k$ and $\mathbf{W}_v$ are trainable parameter matrices with dimensions of $d\times d$, $2d\times d$ and $2d\times d$ respectively.

To aggregate features from the local neighborhood users to capture their commonalities for better user representation, we then use GraphSAGE~\citep{hamilton2018inductive} as follows:
\begin{equation}
    \mathbf{X}_{u}^{(2)} = \text{GraphSAGE}(\mathbf{X}_{u}^{(1)}, \mathbf{A}_u),
\end{equation}
where $\mathbf{X}_{u}^{(2)}$ is the updated user representations containing user interaction information.

In our model, we employ DiffPool~\citep{ying2019hierarchical} for graph pooling on top of $\mathbf{X}_u^{(2)}$. DiffPool is used to identify latent clusters of like-minded users who may share similar views and vulnerabilities. These clusters enhance our ability to learn and perform specific tasks more effectively.
Specifically, DiffPool is used to coarsen the graph $\mathcal{G}_u$ from $|\mathbf{U}|$ user nodes to $|\mathbf{V}_c|$ community nodes. This is done in two steps of standard DiffPool: 1) compute a soft community assignment matrix $\mathbf{C} \in \mathbb{R}^{|\mathbf{U}|\times |\mathbf{V}_c|}$ using GraphSAGE again as below:
\begin{equation}
   \mathbf{C} = \text{softmax}(\text{GraphSAGE}(\mathbf{X}_{u}^{(2)},\mathbf{A}_u)),
    \label{e:s}
\end{equation}
and 2) get the community node embedding matrix $\mathbf{X}_c =\mathbf{C} \mathbf{X}_{u}^{(2)}$ for the new coarsened graph.

Features of both original user graph and pooled community graph are then fed into output layers for rumor detection, virality and vulnerability prediction tasks, as shown in Figure~\ref{framework1}(c).

\subsection{Output Layers}
\label{tsl}
\subsubsection{Graph Classification for Rumor Detection and Virality Prediction} 
For rumor detection and virality prediction, we try different pooling methods over the pooled graph including sum, mean, max, and an additional DiffPool to further pool the user community graph into a single graph representation. We find sum pooling performs best, which might be because the sum operation is more sensitive to the size of the graph.
After obtaining the final graph representation, we use multi-layer perceptrons (MLPs) for rumor detection and virality prediction at the graph level.

\subsubsection{Node Classification for User Vulnerability Prediction}
Community information is expected to help improve the completeness of user information because users within a community often have similar behaviors and interests. 
For vulnerability prediction, we propose a method named \textbf{Community-enhanced Vulnerability Propagation} (CVP) to exploit the latent communities to refine user node representation. The basic idea is to concatenate node representation with its corresponding community representation. 

With the soft community assignment matrix $\mathbf{C}$ and community node embedding matrix $\mathbf{X}_c$, we get an embedding matrix $\mathbf{X}_u^{c} = \mathbf{C}\mathbf{X}_c \in \mathbb{R}^{|\mathbf{U}|\times d}$, where each row is the community embedding of the corresponding node in $\mathbf{X}_{u}^{(2)}$.
Then, we get the updated user embedding by concatenating the original node embedding with its corresponding community embedding
$\mathbf{X}_{u}^{(3)}=[\mathbf{X}_{u}^{(2)},\mathbf{X}^{c}_u]$.
Finally, we pass the enhanced node representations through a GraphSAGE again to get the final user representations $\mathbf{X}_{u}^{(4)} = \text{GraphSAGE}(\mathbf{X}_{u}^{(3)}, \mathbf{A}_u)$, so that embedding of neighboring nodes can influence each other explicitly. 

After obtaining the final node representations, we use an MLP to predict the vulnerability of each user.

\subsection{Training Strategies} 
During training, we use the mean squared error (MSE) loss function for user vulnerability prediction and virality prediction, and cross-entropy loss for rumor detection. 
We train the joint learning framework under two different multi-task settings to explore their ability to mitigate training conflict of multiple tasks: 1) concurrent training, and 2) meta-training based on meta-learning.

\subsubsection{Concurrent Training} 

To deal with the potential training conflict between three tasks and further improve the model performance, we combine the three loss functions using Gradnorm~\citep{chen2018gradnorm} which dynamically adjusts their weights so that the gradient magnitudes of tasks are close and the tasks learn at a similar rate.

\subsubsection{Meta-training}
\label{meta}
\begin{algorithm}[t!]
\caption{Our meta-learning-based multi-task algorithm}
\label{alg:algorithm}
\textbf{Input}: A set of graphs $\mathcal{G}$; Parameters $\theta=\{\theta_b,\{\theta_{h_k}\}\}$.

\begin{algorithmic}[1] 
\STATE Initialize parameters $\theta$ randomly.
\FOR{$iter$ from $1$ to $maxIter$}
\STATE $outer\_loss \leftarrow 0$
\FOR{each task ${\mathcal{T}_{k}}~( k\in\{0,1,2\})$}
\STATE $\theta_{h_k}^{\prime} \leftarrow \theta_{h_k}$
\STATE Compute inner loss:
$inner\_loss =  \mathcal{L}_{\mathcal{T}_{k}}\left(f_{\theta}\right)$
\STATE Update task-specific parameters with gradient descent: \\$\theta_{h_k}^{\prime} \leftarrow \theta_{h_k}-\nabla_{\theta_{h_k}} inner\_loss$ \label{step6}
\STATE 
Update outer loss:
$outer\_loss \leftarrow outer\_loss + \mathcal{L}_{\mathcal{T}_{k}}\left(f_{\theta_{b}, \theta_{h_k}^{\prime}}\right)$\label{step7}

\ENDFOR

\STATE $\theta \leftarrow \theta-\nabla_{\theta} outer\_loss$

\ENDFOR
\STATE Return $\theta$
\end{algorithmic}
\end{algorithm}
Inspired by~\citep{buffelli2020meta}, we use a meta-learning strategy to mitigate training conflicts.
Let's denote the proposed method as $f_\theta$, where $\theta$ represents its parameters.
The goal is to find a set of parameters $\theta$ that can perform well on the three tasks with only a few steps of gradient descent on each task. 
Specifically, $\theta$ is partitioned into the backbone parameters $\theta_{b}$ and the head parameters $\{\theta_{h_k}\}$ where the backbone of our model are the input and refined embedding layers, and the three heads correspond to specific tasks. 
As shown in Algorithm~\ref{alg:algorithm}, the meta-learner tries to find the best configuration via inner and outer loops. For each task, the inner loop updates the head layer parameters $\{\theta_{h_k}\}$ in step~\ref{step6} through a few steps of gradient descent on a task-specific inner loss $\mathcal{L}_{\mathcal{T}_{k}}\left(f_{\theta}\right)$. After updating head parameters, each task-specific loss $\mathcal{L}_{\mathcal{T}_{k}}\left(f_{\theta_{b}, \theta_{h_k}^{\prime}}\right)$ is used to compute the outer loss.
\section{Experimental Evaluation}\label{sect:experiment}

\subsection{Datasets and Setup}\label{sect:dataset}
\begin{table*}[t!]
\centering
\caption{Datasets statistics.}
\begin{tabular}{ccccccc}
\toprule
\multirow{2}*{\textbf{Dataset}}&\multirow{2}*{\textbf{Type}}&\multirow{2}*{\textbf{\# instances}}&\multirow{2}*{\textbf{\shortstack[c]{\textbf{Average} \\ \textbf{\# posts}}}}&\multirow{2}*{\shortstack[c]{\textbf{Average} \\ \textbf{\# users}}}&\multirow{2}*{\textbf{\shortstack[c]{\textbf{Average} \\ \textbf{Vulnerability}}}}&\multirow{2}*{\textbf{\shortstack[c]{\textbf{Average} \\ \textbf{Virality}}}}\\\\\midrule
\multirow{2}*{\textbf{TWITTER}} &\textbf{Rumor}&1,560&277.2&271.7&0.843&340.1\\
&\textbf{Non-Rumor}&579&503.7&498.3&0.196&621.2 \\\midrule
 \multirow{2}*{\textbf{WEIBO}}&\textbf{Rumor}&2,311&720.7&701.2&0.904&876.4 \\
&\textbf{Non-Rumor}&2,351&577.9&568.7&0.124&711.0\\\bottomrule

\end{tabular}
\label{vir_sta}

\end{table*}

\subsubsection{Datasets}
We construct our data based on two public datasets namely TWITTER~\citep{ma2017detect} and WEIBO~\citep{ma2016detecting}. The original datasets were designed for rumor detection with only rumor/non-rumor annotations at the graph level. We need to derive ground-truth virality labels and vulnerability labels. 
We establish the \emph{virality} label for each graph based on the count of unique users in the entire propagation process of each event in our datasets.
Our goal is to predict the virality of a propagation network using its earlier propagation state as input (e.g. when the observation percentage of the propagation $\frac{t}{T}$ equals to 20\%, 40\%, ..., etc.). Following~\cite{liao2019popularity}, our main tables (Table~\ref{tbl:results},~\ref{tbl:ab_results},~\ref{tbl:task_ab_results} and~\ref{tbl:neg_trans}) show the prediction results obtained when $\frac{t}{T}=80\%$. 
The prediction results corresponding to other proportions as observations are provided in Figure~\ref{fig:final}.
For users, we use the proportion of rumors among all events in which the user is involved as their gold \emph{vulnerability}.
Only users engaged in more than one propagation graph are labeled, constituting 15.5\% users on TWITTER and 16.7\% users on WEIBO. Unlabeled users are not considered in performance evaluation.

We split the dataset into training, validation, and test sets by $80\%, 10\%$, and $10\%$, respectively. A key priority is
to eliminate user overlaps between the training set and the validation/test sets, thereby ensuring no data leakage. To accomplish this, we began by examining each graph within our datasets. If a graph exclusively contains users not found
in any other graphs, it is designated as a \emph{non-overlapping} graph.
Notably, over 20\% of the graphs in our datasets are non-overlapping with any other graphs. We then randomly sample from these non-overlapping graphs, so that both the validation and test sets comprise these sampled non-overlapping graphs, with each of the two sets representing 10\% of the entire dataset. Finally, the rest of the non-overlapping graphs and all other graphs are put in the training set. This ensures that users from the users in the validation/test sets remain unseen during the training process. Table~\ref{vir_sta} gives the statistics of our datasets. 

\subsubsection{Metrics}
For the rumor detection task, since the accuracy of the classification results can be affected by the class imbalance, we further use precision, recall, and macro-averaged F1 score (MacF1) score in addition to accuracy to provide a more comprehensive evaluation of performance. 
Meanwhile, for the virality prediction and vulnerability prediction tasks, we use MSE, and mean squared logarithmic error (MSLE) to evaluate the regression results. In addition, we also use normalized Discounted Cumulative Gain (nDCG)~\citep{jarvelin2000ir} to measure the models' ranking performance based on the predicted virality and vulnerability scores because nDCG takes into account the relevance and ranking position of each predicted item, which is particularly important for virality and vulnerability prediction tasks where the order of importance of the predicted items can significantly affect the overall effectiveness of the model.

\subsubsection{Parameter Settings} 
\begin{figure}
\centering
\subfigure[Number of Network Layers]{
\includegraphics[width=.3\linewidth]{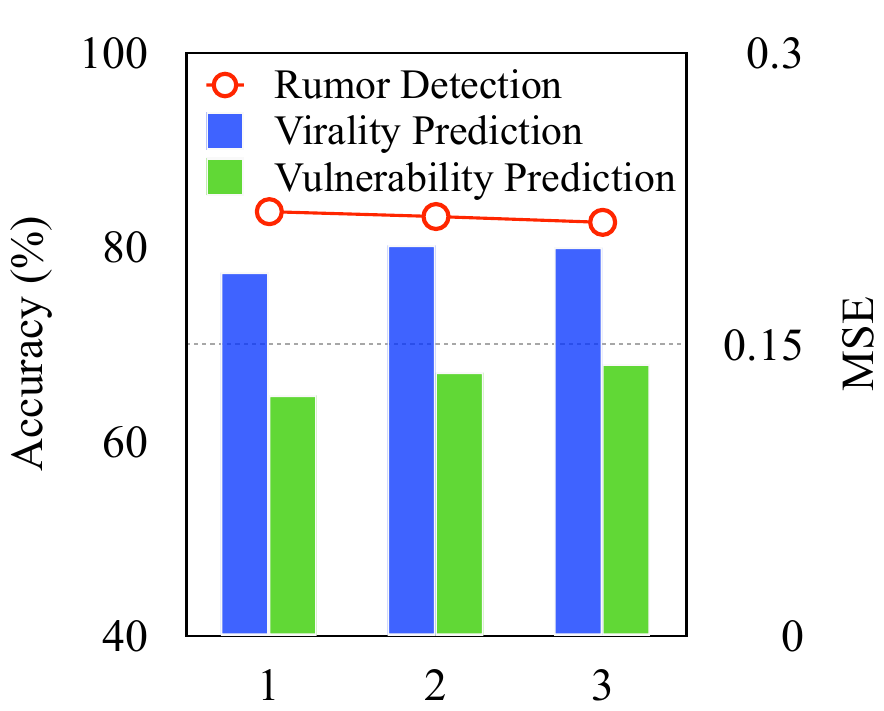}
}
\subfigure[Embedding Dimension]{
\includegraphics[width=.3\linewidth]{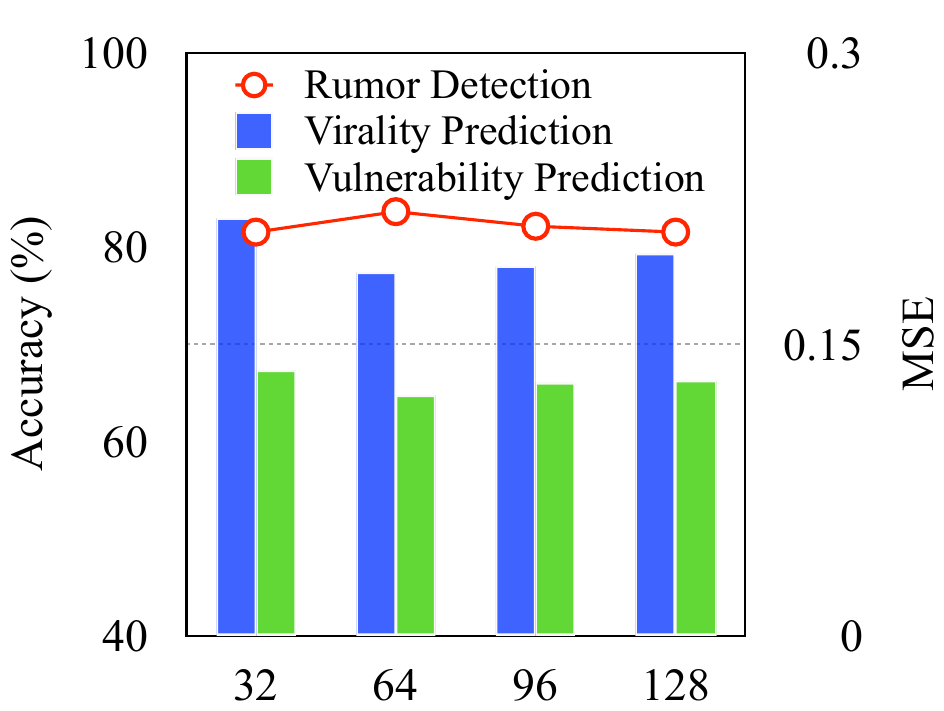} 
}
\subfigure[Maximum Post Length]{
\includegraphics[width=.3\linewidth]{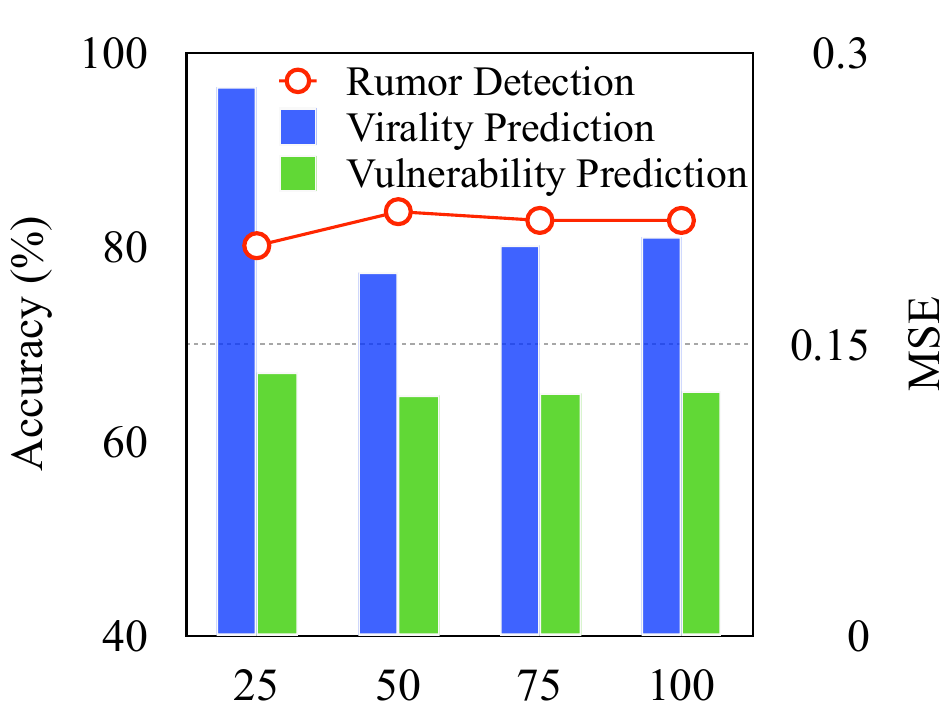} 
}

\caption{The impact of hyper-parameters that are related to model's complexity based on the validate set.}
\label{fig:hyper}
\end{figure}

We implement all GNN-based models with DGL\footnote{\url{https://www.dgl.ai/}}, and all deep learning based models with Pytorch\footnote{\url{https://pytorch.org/}}. Our method is based on GraphSAGE~\citep{hamilton2018inductive} and DiffPool~\citep{ying2019hierarchical} models.
In our experiments, we employ Random Search~\citep{bergstra2012random} for hyper-parameter tuning since it is more efficient than Grid Search. We set the number of iterations in Random Search as 100. Our focus was on a number of key hyper-parameters: 1) the number of layers in both GraphSAGE and CVP, drawn from $\{1, 2, 3\}$; 2) the pool size for DiffPool, adjusted with the choices in $\{25, 50, 75, 100\}$; 3) the embedding dimension, selected from $\{32, 64, 96, 128\}$; 4) batch size, selected from $\{2,4,6,8,10\}$; 5) the learning rate, selected from the set $\{1e-2,5e-3, 1e-3\}$; 6) the dropout rate, sampled from the set $\{0.1, 0.2, 0.4, 0.6\}$; and 7) the maximum post length, selected from the set $\{25, 50, 75, 100\}$.
We examine the influence of these hyper-parameters on the validate set. In particular, three of them directly determine the model's complexity, i.e., the number of layers, the embedding dimension, and the maximum post length. We specifically demonstrate their influence on model performance in Figure~\ref{fig:hyper}. Random Search led us to select one layer for both GraphSAGE and CVP, a DiffPool pool size of 50, an embedding dimension of 64, a batch size of 8, a learning rate of 5$e$-3, a dropout rate of 0.2, and a maximum post length of 50.

During training, we update model parameters using stochastic gradient descent and optimize the model by Adam algorithm. The experimental results are averaged over five independent runs with different seeds.

\subsection{Results of Rumor Detection.}

\begin{table*}[t!]
\begin{subtable}[Rumor Detection]{
\centering
\begin{tabular}{lllllllllll}
\toprule
\multirow{3}*{\textbf{\large Model}}& \multicolumn{4}{c}{\textbf{TWITTER}}& \multicolumn{4}{c}{\textbf{WEIBO}}\\ 
\cmidrule(lr){2-5} \cmidrule(lr){6-9}
& Acc. & Pre. &Recall  & MacF1        & Acc.     & Pre. &Recall& MacF1   \\\cmidrule(lr){1-1}\cmidrule(lr){2-5} \cmidrule(lr){6-9}
GCNFN&0.772&0.664&0.714&0.731&0.902&0.885&0.930&0.908\\
Bi-GCN&0.790&0.736&0.763&0.716&0.913&0.893&0.942&0.921\\
RoBERTa&0.791&0.749&0.772&0.740&0.917&0.907&0.940&0.923\\
UPFD&0.815&0.836&0.783&0.805&0.921&0.905&0.940&0.933\\
DDGCN&0.813&0.821&0.790&0.811&0.918&0.911&0.937&0.925\\
Us-DeFake&0.819&0.839&0.803&0.821&0.919&0.910&0.931&0.924\\
MT-CON (ours)& 0.820&0.854$^*$&0.814$^*$&0.833&0.946$^*$&0.929$^*$&0.949$^*$&0.943$^*$\\
MT-META (ours)&\textBF{0.826}$^*$&\textBF{0.879}$^{**}$&\textBF{0.826}$^*$&\textBF{0.845}$^*$&\textBF{0.954}$^*$&\textBF{0.941}$^{**}$&\textBF{0.955}$^*$&\textBF{0.952}$^{**}$\\
\bottomrule
\end{tabular}}
\end{subtable}
\hfill
\begin{subtable}[Virality Prediction]{
\centering
\begin{tabular}{lllllll}
\toprule
\multirow{3}*{\textbf{\large Model}}& \multicolumn{3}{c}{\textbf{TWITTER}}& \multicolumn{3}{c}{\textbf{WEIBO}}\\ 
\cmidrule(lr){2-4} \cmidrule(lr){5-7}
& MSE    & MSLE   & nDCG    & MSE    & MSLE    & nDCG       \\ \cmidrule(lr){2-4} \cmidrule(lr){5-7}
DeepHawkes&2.011&0.067&0.521&0.994&0.048&0.783\\
NPP&1.199&0.020& 0.867&0.867&0.022&0.984\\
DeepBlue& 0.918&0.016&0.894&0.824&0.017&0.985\\
BERT& 0.960& 0.019&0.891&0.835&0.018&0.989\\
CasSeqGCN&0.648&0.009&0.990&0.799&0.013&0.988\\
TCAN&0.588&0.008&0.992&0.792&0.017&0.989\\
MT-CON (ours) &0.320$^{**}$&0.006&0.996&0.716$^{**}$&0.013&\textBF{0.995}\\
MT-META (ours) &\textBF{0.197}$^{**}$&\textBF{0.005}&\textBF{0.999}&\textBF{0.603}$^{**}$&\textBF{0.008}$^{**}$&\textBF{0.998}$^*$ \\
\bottomrule
\end{tabular}}
\end{subtable}
\hfill
\begin{subtable}[Vulnerability Prediction]{
\centering
\begin{tabular}{lllllll}
\toprule
\multirow{3}*{\textbf{\large Model}}& \multicolumn{3}{c}{\textbf{TWITTER}}& \multicolumn{3}{c}{\textbf{WEIBO}}\\ 
\cmidrule(lr){2-4} \cmidrule(lr){5-7}
& MSE    & MSLE   & nDCG    & MSE    & MSLE    & nDCG       \\ \cmidrule(lr){2-4} \cmidrule(lr){5-7}
LING-GAT&0.121&0.062&0.989&0.168&0.074&0.972\\
GraphRfi&0.151&0.075&0.985&0.179&0.079&0.963\\
IMP-GCN&0.140&0.074&0.975&0.181&0.079&0.964\\
U-BERT&0.124&0.065&0.983&0.166&0.084&0.974\\
PinnerFormer&0.126&0.067&0.985&0.167&0.082&0.969\\
CLUE&0.127&0.068&0.987&0.164&0.077&0.972\\
MT-CON (ours)&0.112&0.054&0.991&0.153&0.061$^{*}$&0.981\\
MT-META(ours)&\textBF{0.104}$^*$&\textBF{0.035}$^{**}$&\textBF{0.995}&\textBF{0.137}$^{*}$&\textBF{0.047}$^{**}$&\textBF{0.988}$^*$\\
\bottomrule
\end{tabular}}
\end{subtable}
\caption{Experiment results. $*$ ($**$): Significant improvement over the best baseline with $p<0.05$ ($0.01$). Bold denotes the best performance in each task.} 

\label{tbl:results}
\end{table*}

Given our multi-task models based on concurrent training (MT-CON) and meta-learning (MT-META), we make comparisons with the following baseline models:
\begin{itemize}
    \item GCNFN~\citep{monti2019fake}: A rumor detection model by exploiting geometric deep learning.
    \item Bi-GCN~\citep{bian2020rumor}: A rumor detection method based on bidirectional GCN.
    \item RoBERTa~\citep{pelrine2021surprising}:  A baseline directly fine-tuning the pre-trained RoBERTa Model, which achieves surprising performance on rumor detection tasks.
    \item UPFD~\citep{dou2021user}: A GCN-based model that considers user preference to help rumor detection.
    \item DDGCN~\citep{sun2022ddgcn}: A Dual-Dynamic GCN based model, which can model the dynamics of propagation networks as well as the dynamics of the background knowledge from knowledge graphs.
    \item Us-DeFake~\citep{su2023mining}: A rumor detection method by learning the propagation features and the user interaction features.
\end{itemize}
As shown in Table~\ref{tbl:results}, our proposed joint learning models generally outperform all the baselines on both datasets. We have the following observations:

(1) \textit{GCN with proper design is one promising way in rumor detection.} GCN-based models can effectively identify the propagation patterns of rumors and non-rumors, and are widely used as base models in rumor detection. The integration of prior knowledge through special design can further enhance the models' performance in this task. For example, by taking user preference characteristics into account, 
On average, UPFD surpasses GCNFN and Bi-GCN in terms of accuracy by 3.8\% and 2.0\%, respectively.

(2) \textit{Content in posts and their responses provides valuable information for detecting rumors.} Although ignoring the complex graph structures and only exploiting the content information, fine-tuning the pre-trained RoBERTa model can also achieve comparable behavior with GCN-based GCNFN and Bi-GCN models. For example, RoBERTa surpasses GCNFN in terms of accuracy by 2.1\%. This suggests the importance of content information unleashed by the strong presentation ability of the pre-trained model for rumor detection.

(3) \textit{Hierarchical graph pooling can help utilize more useful information.} Node representations learned from both basic GCNs are flat since they cannot encode the graph structure in a hierarchical way by just propagating information through nodes and edges. Our proposed models MT-CON and MT-META take advantage of user representation and hierarchical pooling in joint learning, thus beating all the baselines. For example, MT-META surpasses the best baseline Us-DeFake in terms of accuracy by 2.4\% on average.

(4) \textit{Mitigating training conflicts of multiple tasks can further improve performance.} MT-META outperforms MT-CON in terms of accuracy by 0.8\% on average, suggesting the meta-learning strategy behaves better in mitigating training conflicts for our tasks compared to concurrent training.

\subsection{Results of Virality Prediction.}

We make comparisons with the following baseline models:
\begin{itemize}
    \item DeepHawkes~\citep{10.1145/3132847.3132973}: A Hawkes process-based deep learning model for virality prediction.
    \item NPP~\citep{chen2019npp}: A prediction model that learns embeddings for popularity and virality, taking into account time, user, and content factors.
    \item DeepBlue~\citep{zhang2021deepblue}: A popularity and virality method based on bi-layered LSTM, which takes into account historical information such as user reputation and tweet-related features.
    \item BERT~\citep{tan2022efficient}: A Bert-based model to exploit content features for popularity and virality prediction. The original paper exploits features from other modalities (e.g. images). Here we only use its text processing method for comparison.
    \item CasSeqGCN~\citep{wang2022casseqgcn}: A popularity and virality prediction method that employs GCN for network structure features and LSTM for temporal dynamics.
    \item TCAN~\citep{sun2023explicit}: An explicit time embedding based popularity and virality method, which employs a  graph attention encoder and a sequence attention encoder to learn the representation of propagation networks.
\end{itemize}

As shown in Table~\ref{tbl:results}, MT-CON and MT-Meta generally outperform all the baselines on both datasets. We have the following observations:

(1) \textit{Information contained in propagation structures can aid in virality prediction but may not be sufficient.} DeepHawkes performs the worst, with 39.2\% higher in MSE than the second worst model (i.e., NPP), as it only uses the network structure to model the discussion process. Models considering more information like context and user perform clearly better.

(2) \textit{Incorporating user and event features can improve performance in virality prediction.} DeepBlue outperforms NPP by 13.3\% lower in MSE on average, as it considers not only individual tweets but also incorporates useful user reputation and post-related features learned from historical tweets. Additionally, our model considers the information type of event (i.e., rumor/non-rumor) and the vulnerability of the users involved in its propagation to aid in virality prediction.

(3) \textit{Content in posts and their responses can signal the future virality of the current event.} Although only exploiting content features, the pre-trained BERT model achieves comparable behavior to DeepBlue, showing only a slight increase in MSE by 3.3\%. This suggests the significant role of text in the virality prediction task as well as the strengths of the pre-trained model.

(4) \textit{Incorporating hierarchical pooling can lead to a better understanding of the graph's structure, thus improving the accuracy of virality prediction.} Our models MT-CON and MT-META utilize a combination of different feature representations such as time, user, and post content and information on event type and user vulnerability, which are then hierarchically pooled, leading to improved performance compared to all baselines. For example, MT-META exhibits a lower MSE than TCAN by 49.3\% on average.

(5) \textit{Reducing training conflicts between multiple tasks can enhance performance.} Similar to the observation in rumor detection results, MT-META performs better than MT-CON in mitigating training conflicts with the help of a meta learning-based training strategy. On average, MT-META shows a lower MSE than MT-CON by 37.5\%.

\subsection{Result of User Vulnerability Prediction}

For this task, we make comparisons with the following baseline models:
\begin{itemize}
    \item LING-GAT~\citep{del2019you}: A user representation method  using Bi-LSTM layers to capture linguistic features and a transductive graph attention network (GAT) to model a user's social relationships.
    \item GraphRfi~\citep{zhang2020gcn}: A hand-crafted feature based user representation learning framework using GCN and neural random forest.
    \item IMP-GCN~\citep{liu2021interest}: A GCN-based user interest-aware representation learning model.
    \item U-BERT~\citep{qiu2021u}: A pre-trained user embedding model inspired by the success of the BERT model.
    \item PinnerFormer~\citep{pancha2022pinnerformer}:
    A user representation learning method based on a sequence of users' recent actions. To make it adapt to the vulnerability prediction task, we treat users' posts and comments as their actions.
    \item CLUE~\citep{shin2023scaling}: A contrastive learning based user representation learning method for general purpose.
    
\end{itemize}
To adapt existing baseline models for the regression task of predicting user vulnerability, we modified them by removing their task-specific final layers and replacing them with a regression layer. This allowed us to use the same model architectures while repurposing them for the specific task of user vulnerability prediction.

As shown in Table~\ref{tbl:results}, our proposed models outperform all the baselines on both datasets for vulnerability prediction. We have some specific observations:

(1) \textit{Hand-crafted features are limited in capturing latent features and deep correlations.} GraphRfi performs the worst with 7.1\% higher in MSE than the second worst model (i.e., IMP-GCN), due to the generally weak generalizability of hand-crafted features. 

(2) \textit{The content of user posts can reveal their inner vulnerability.} IMP-GCN performs slightly better than GraphRfi by 6.6\% lower in MSE, as it uses content information that can convey user propensity. However, it still performs worse than another GCN-based model LING-GAT by 16.5\% higher in MSE, as LING-GAT deeply exploits post content information with Bi-LSTM layers.

(3) \textit{Pre-training and contrastive learning are promising ways in user representation learning.} U-BERT outperforms all other baselines, suggesting the usefulness of pre-trained models for user classification respectively. Interestingly, PinnerFormer and CLUE, which do not utilize propagation network structures, demonstrate impressively good performance. For instance, CLUE achieves an average MSE that is 12.6\% and 10.7\% lower than GraphRfi and IMP-GCN, respectively, which consider more information. This points to the untapped potential of contrastive learning, a technique employed by both PinnerFormer and CLUE. Our model also pre-trains a general user embedding via contrastive learning to help capture user features.

(4) \textit{Decreasing training conflict of multiple tasks leads to improved performance.} This observation is similar to the other two tasks. In the vulnerability prediction task, MT-META achieves a lower MSE than MT-CON by 8.9\% on average.

\subsection{Analysis}

\subsubsection{Ablation Study}
To examine the impact of each key component in our proposed joint learning approach, we perform an ablation study based on our best full model MT-META on the TWITTER dataset. Table~\ref{tbl:ab_results} shows our experiments. We summarize various aspects and highlight the most interesting findings below.
\paragraph{Input Embedding.} We first show the effect of each component in the input embedding layer in the first four rows. We can see that there is a moderate degradation in the model performance when the pre-trained user embeddings (i.e., User Emb. in Table~\ref{tbl:ab_results}) are not used, which suggests the importance of mining deep structure and patterns hidden in the constructed global user graph for all our three tasks.
In addition, in our model, both content and time information in post embedding play indispensable roles, as demonstrated by the performance drop after the removal of content, time, and both types of embeddings. 
\paragraph{Refined Embedding.}
If we replace hierarchical pooling with sum pooling, which means that we directly sum over the node representations to get the graph representations, the performance drops. This is consistent with our hypothesis that learning the hierarchical structure of the graphs (i.e., through DiffPool in our model) can improve the model's performance.

\paragraph{CVP}
Both community enhancement and GraphSAGE in CVP play important roles in user vulnerability prediction. This further confirms our hypothesis that CVP can directly help predict user vulnerability by utilizing the shared features among users within latent communities to refine node representation.

\begin{table*}[thb!]
\centering
\small
\caption{Ablation study results on TWITTER.} 
\begin{tabular}{c|ccc|cc|cc||cccccc}
\toprule
\multirow{3}*{\#}&\multicolumn{3}{c}{\textbf{Input Emb.}}& \multicolumn{2}{c}{\textbf{Refined Emb.}}&\multicolumn{2}{c}{\textbf{CVP}}&\multicolumn{6}{c}{\textbf{Task}}\\
\cmidrule(lr){2-4}\cmidrule(lr){5-6}\cmidrule(lr){7-8}\cmidrule(lr){9-14}
&\multirow{2}*{\shortstack[c]{User\\Emb.}}&\multirow{2}*{\shortstack[c]{Cnt.\\Emb.}}&\multirow{2}*{\shortstack[c]{Time \\ Emb.}}&\multirow{2}*{\shortstack[c]{Sum \\ Pool}}&\multirow{2}*{\shortstack[c]{Hier. \\ Pool}}&\multirow{2}*{\shortstack[c]{Comm.\\Enh.}}&\multirow{2}*{\shortstack[c]{Graph- \\ SAGE}}& \multicolumn{2}{c}{\textbf{Rumor}}& \multicolumn{2}{c}{\textbf{Virality}}& \multicolumn{2}{c}{\textbf{Vulnerability}}\\ 
\cmidrule(lr){9-10}\cmidrule(lr){11-12}\cmidrule(lr){13-14}
  &&&&&&&  &Acc&MacF1& MSE& nDCG& MSE& nDCG\\\midrule
1&&\checkmark&\checkmark&&\checkmark&\checkmark&\checkmark&0.802&0.757&0.233&0.995&0.138&0.983 \\
2&\checkmark&&&&\checkmark&\checkmark&\checkmark&0.777&0.742&0.391&0.990&0.159&0.988\\
3&\checkmark&&\checkmark&&\checkmark&\checkmark&\checkmark&0.784&0.759&0.373&0.990&0.152&0.986\\
4&\checkmark&\checkmark&&&\checkmark&\checkmark&\checkmark&0.791&0.771&0.381&0.992&0.143&0.984\\\midrule

5&\checkmark&\checkmark&\checkmark&\checkmark&&\checkmark&\checkmark&0.777&0.730&0.405&0.985&0.160&0.984  \\\midrule 
6&\checkmark&\checkmark&\checkmark&&\checkmark&&&0.821&0.817&0.211&0.994&0.152&0.986\\
7&\checkmark&\checkmark&\checkmark&&\checkmark&&\checkmark&0.821&0.809&0.204&0.998&0.147&0.979\\
8&\checkmark&\checkmark&\checkmark&&\checkmark&\checkmark&&0.822&0.814&0.204&0.998&0.134&0.984\\\midrule
9&\checkmark&\checkmark&\checkmark&&\checkmark&\checkmark&\checkmark&\textBF{0.826}&\textBF{0.845}&\textBF{0.197}&\textBF{0.999}&\textBF{0.104}&\textBF{0.995}\\
\bottomrule
\end{tabular}
\label{tbl:ab_results}
\end{table*}

\subsubsection{Effects of Different Task Loss Configurations} 
To assess the contribution of each task loss in our proposed joint learning method, we conduct an ablation experiment by varying the loss configuration and using our most effective model MT-META on the TWITTER dataset, as shown in Table~\ref{tbl:task_ab_results}. We have some specific observations:
\begin{itemize}
    \item \textit{The direct connection between virality and rumorous class (i.e., rumor or non-rumor) is weak.} When only the virality loss is taken into account, the performance of rumor detection is the lowest. Similarly, the performance of virality prediction is the lowest when only the rumor detection loss is considered.
    \item \textit{Task training conflict can cause performance degradation.}
    Inappropriate joint learning may result in a certain level of training conflict, as it attempts to balance the learning process among different task objectives. 
    Take the rumor detection task as an example, when the rumor detection loss is jointly optimized with the virality prediction loss, the performance drops by 5.1\% in accuracy and 0.6\% in MacF1 than only the rumor detection loss being optimized; when jointly optimized with the vulnerability prediction loss, the accuracy slightly decreases from 0.806 to 0.801 while MacF1 slightly increases from 0.796 to 0.8. This might suggest that rumor prediction is more directly related to user vulnerability than virality.
    \item \textit{Joint learning across all three tasks is essential to prevent training conflict.} This is because given user vulnerability as a bridge, the relationship between the virality and the rumor/non-rumor nature of information could be established implicitly.
    In the rumor detection task, we observe a 2.5\% and 6.2\% increase in accuracy and MacF1 respectively in the rumor detection task when jointly optimizing all three losses compared to only rumor detection loss. A similar extent of improvement can be observed in the other two tasks. This implies that joint learning across all three tasks can prevent training conflict, probably due to the bridge effect with more tasks. 
\end{itemize}
\begin{table*}[thb!]
\centering
\small
\caption{Effect of different task loss configurations on TWITTER. Underline denotes the performance of a specific task when its corresponding loss was optimized during training. Bold denotes the best performance in each task.}
\begin{tabular}{cccccccccc}
\toprule
\multirow{3}*{\#}&\multicolumn{3}{c}{\textbf{Loss}}&\multicolumn{6}{c}{\textbf{Task}}\\
\cmidrule(lr){2-4}\cmidrule(lr){5-10}
&\multirow{2}*{\shortstack[c]{Rumor\\Detection}}&\multirow{2}*{\shortstack[c]{Virality\\Prediction}}&\multirow{2}*{\shortstack[c]{Vulnerability \\ Prediction}}& \multicolumn{2}{c}{\textbf{Rumor}}& \multicolumn{2}{c}{\textbf{Virality}}& \multicolumn{2}{c}{\textbf{Vulnerability}}\\ 
\cmidrule(lr){5-6}\cmidrule(lr){7-8}\cmidrule(lr){9-10}

  &&&  &Acc&MacF1& MSE& nDCG& MSE& nDCG\\\midrule
1&\checkmark&&&\ul{0.806}&\ul{0.796}&321.766&0.145&23.262&0.005\\
2&&\checkmark&&0.673&0.641&\ul{0.209}&\ul{0.990}&1.977&0.187\\
3&&&\checkmark&0.739&0.704&168.336&0.275&\ul{0.122}&\ul{0.972}\\\midrule
4&&\checkmark&\checkmark&0.748&0.710&\ul{0.201}&\ul{0.998}&\ul{0.137}&\ul{0.961}\\
5&\checkmark&&\checkmark&\ul{0.801}&\ul{0.800}&9.513&0.479&\ul{0.207}&\ul{0.954}\\
6&\checkmark&\checkmark&&\ul{0.765}&\ul{0.742}&\ul{0.249}&\ul{0.990}&0.518&0.872\\\midrule
7&\checkmark&\checkmark&\checkmark&\textBF{\ul{0.826}}&\textBF{\ul{0.845}}&\textBF{\ul{0.197}}&\textBF{\ul{0.999}}&\textBF{\ul{0.104}}&\textBF{\ul{0.995}}\\
\bottomrule
\end{tabular}
\label{tbl:task_ab_results}
\end{table*}
\begin{figure*}
\begin{subfigure}{}
  \centering
\includegraphics[height=3cm]{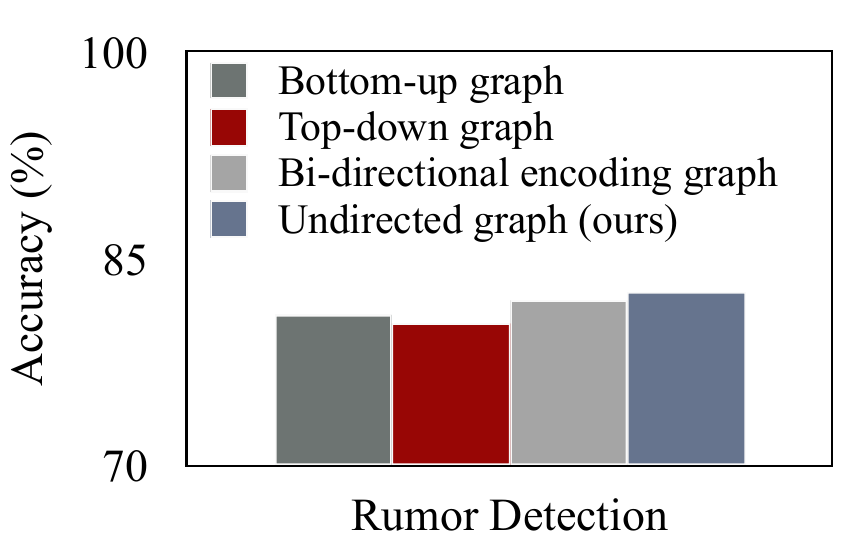}  
\includegraphics[height=3cm]{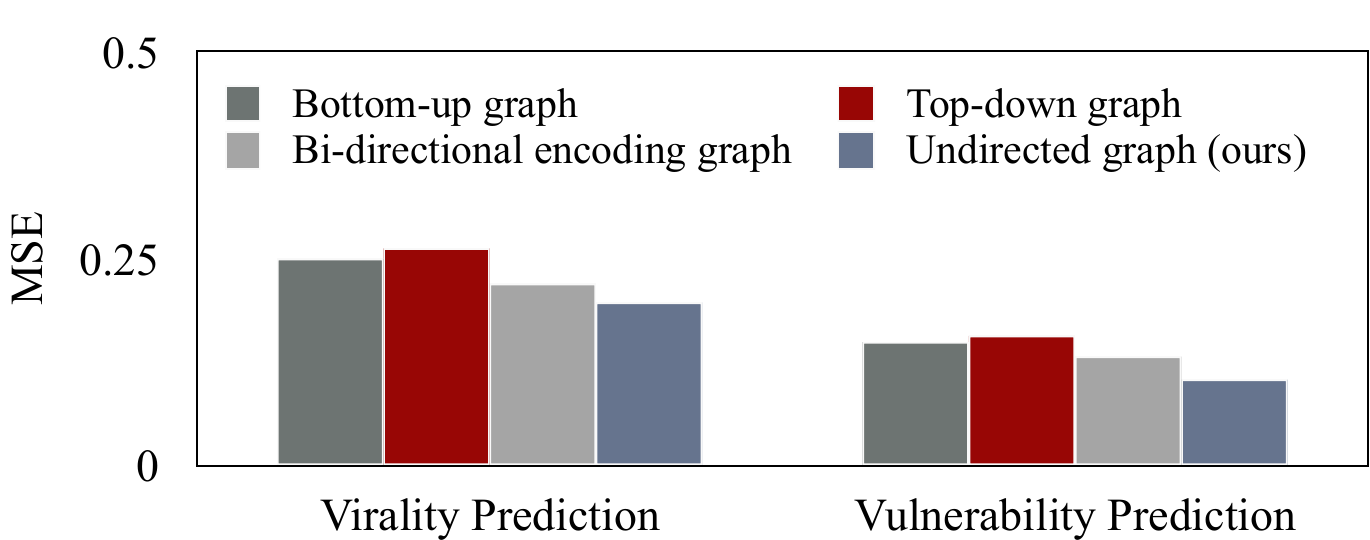}  

\end{subfigure}
\caption{Effect of edge direction.}
\label{fig:edge-direct}
\end{figure*}

\subsubsection{Effects of the Edge Direction}
To investigate the impact of edge direction in propagation networks in our proposed method, we explore four distinct edge direction settings: (1) A bottom-up graph where edges follows the direction of information being referenced~\citep{ma-etal-2018-rumor,bian2020rumor}, (2) A top-down directed graph, where edges follow the direction of information flow~\citep{ma-etal-2018-rumor,bian2020rumor}, (3) An bi-directional encoding
graph, where two GraphSage encoders work simultaneously to represent the features along with the two different directions (i.e., top-down and bottom-up) in the graph. The embeddings produced by these encoders are concatenated to form a comprehensive node representation containing bi-directional information, similar as Bi-GCN~\citep{bian2020rumor}, and (4) Our undirected graph, where each edge, although without direction, indicates a two-way relationship and the adjacency matrix of the graph unequivocally represents
a bidirectional graph as each edge is reciprocated. 
As shown in Figure~\ref{fig:edge-direct}, our findings indicate that both top-down and bottom-up directed graphs exhibit worse performance than the two bi-directional variants in (3) and (4). This may suggest that feature representation with the underlying model considering both directions is better than considering only a single
direction regardless of the actual direction of propagation. Moreover, the undirected treatment in (4) outperforms the bi-directional case in (3). This might be because the undirected graph integrates information from both
directions more directly with the symmetric adjacency matrix, rather than a separate learning and then combining them. Such direct integration helps the model better understand the relationships between nodes.

\subsubsection{Effects of the GNN Encoder}
\begin{figure*}
\begin{subfigure}{}
  \centering
\includegraphics[height=3cm]{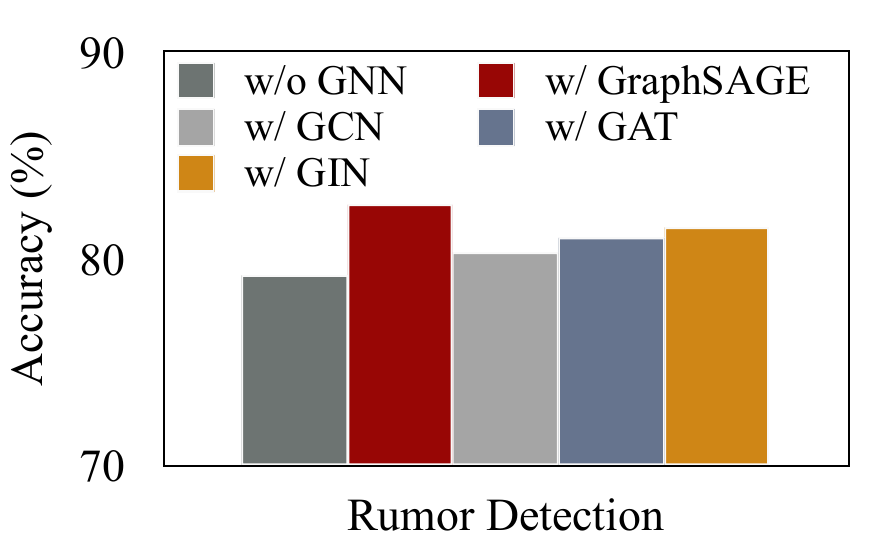}  
\includegraphics[height=3cm]{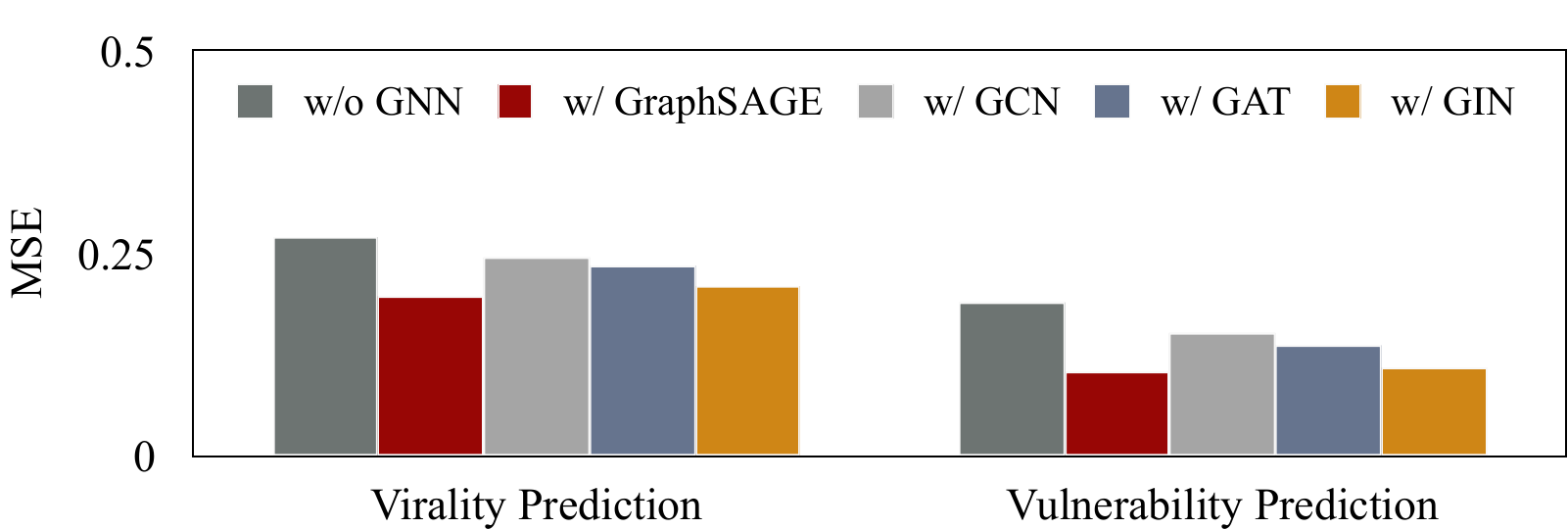}  

\end{subfigure}
\caption{Effect of the GNN Encoder.}
\label{fig:gnn}
\end{figure*}

To investigate the influence of the GNN encoder within our proposed method, we tested different GNN encoders to
derive node representations. In Figure~\ref{fig:gnn}, we display the performance using six distinct encoders, namely GCN~\citep{defferrard2017convolutional},  GraphSAGE~\citep{hamilton2018inductive}, GAT~\citep{2018graph}, GIN~\citep{xu2018powerful}, and a case without a GNN encoder. We observe that the variants with a GNN encoder consistently outperform those without it,
indicating the advantage of GNNs in updating node features. Moreover, we notice that GraphSAGE performs
slightly better than other GNN encoders. We conjecture that this could be due to GraphSAGE’s inductive learning and
sampling mechanism, which might be particularly suitable for social network graphs with abundant noisy information.

\subsubsection{Effects of the Observation Percentage}
\begin{figure}
\centering
\subfigure[Rumor detection]{
\includegraphics[width=.3\linewidth]{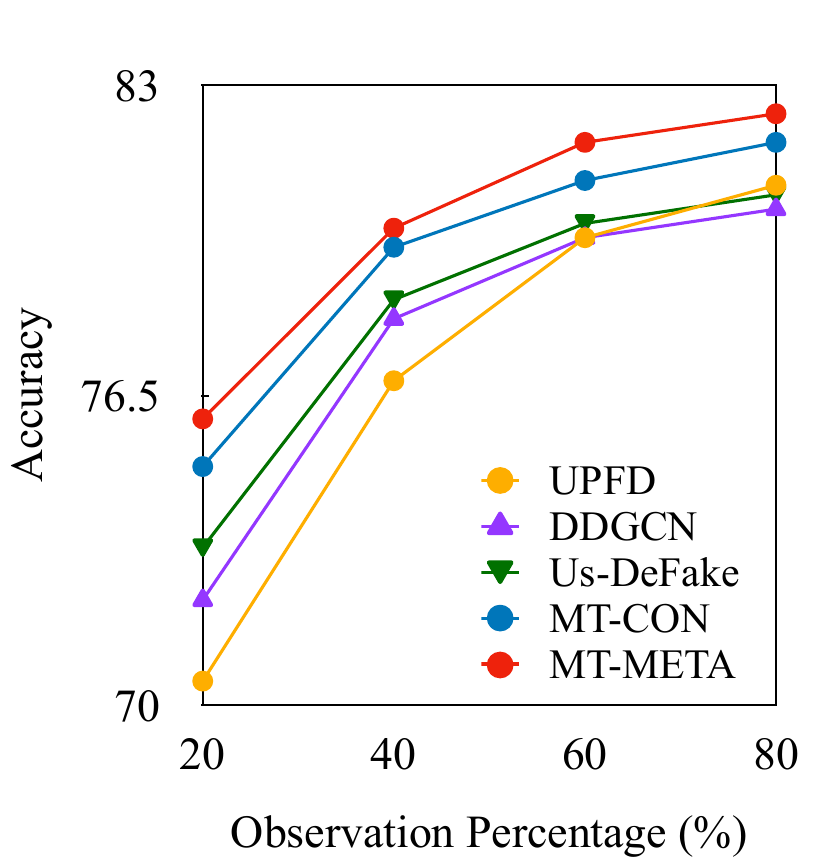}
\label{fig:final_a}
}
\subfigure[Virality Prediction]{
\includegraphics[width=.3\linewidth]{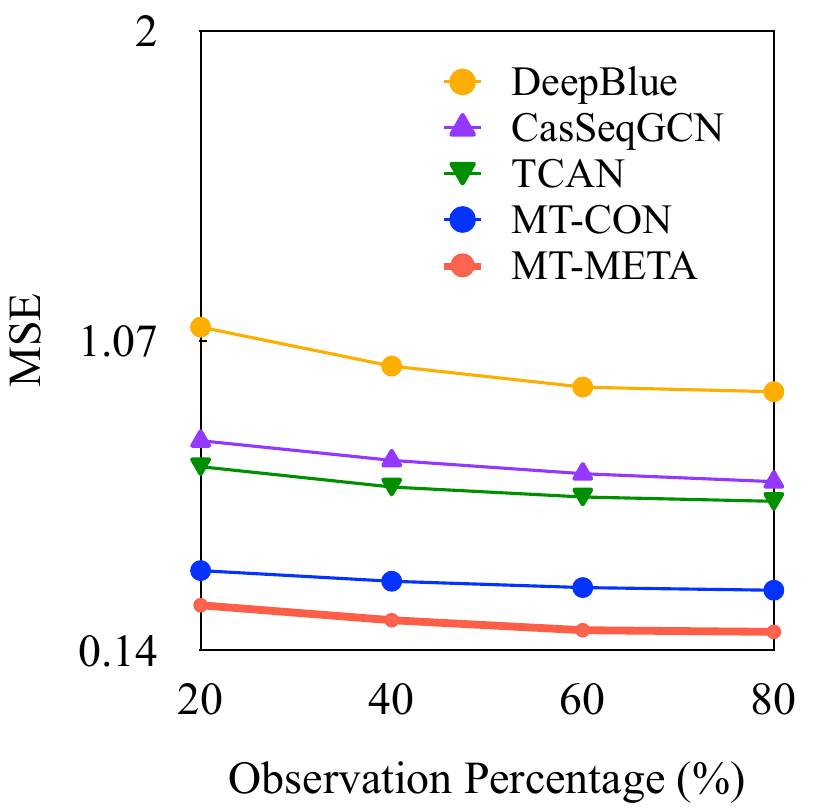} 
}
\subfigure[Vulnerability Prediction]{
\includegraphics[width=.3\linewidth]{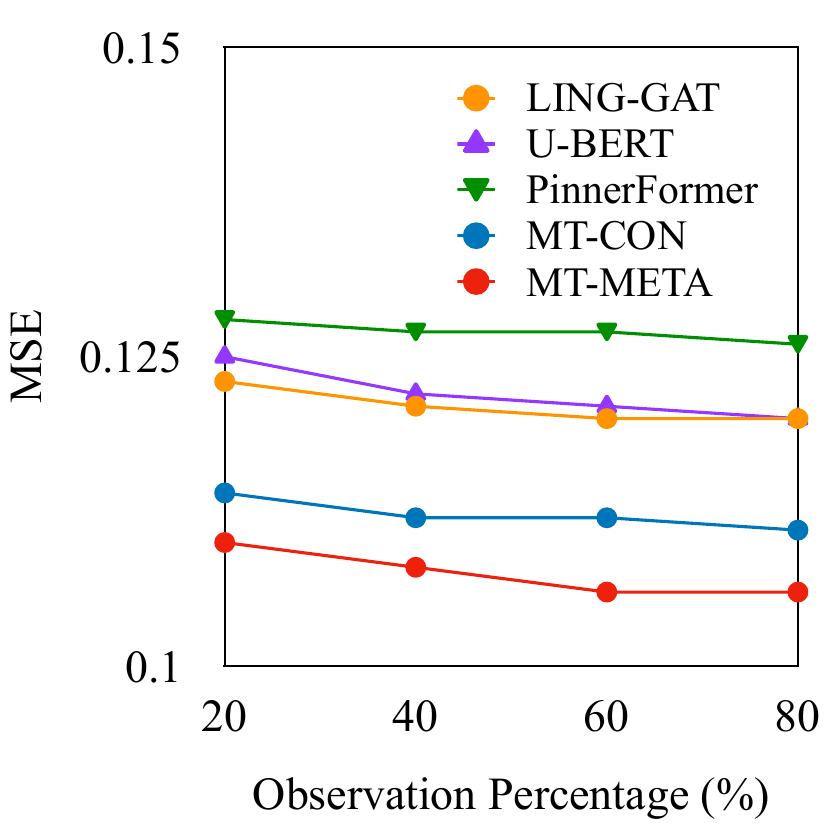} 
}
\caption{Impact of the observed fraction of cascades on the performance of three tasks. }
\label{fig:final}
\end{figure}

For clarity, we only compare our models with the three best-performing baselines for each task. Figure~\ref{fig:final} demonstrates the changes in rumor detection accuracies and the MSEs for virality and vulnerability predictions as the observation percentage $\frac{t}{T}$ changes from 20\% to 80\%. As the percentage increases, the size of the observed graphs also increases, resulting in improved performance in all three tasks. From the figure, we can see that even only using early propagation information (e.g., 20\%), our proposed models MT-CON and MT-META outperform the baseline models with clear margins on the three tasks. In particular, with only 40\% propagation as observation, the MT-META can achieve similar rumor detection accuracy as the three baselines using 80\% propagation (see Figure~\ref{fig:final_a}).

\begin{table}[thb!]
\centering
\caption{Effect of training strategies on the TWITTER dataset. STL/MTL: Single-/Multi-Task Learning.  
} 
\begin{tabular}{cccccccc}
\toprule
\multicolumn{2}{c}{\multirow{2}*{\textbf{Setting}}}& \multicolumn{2}{c}{\textbf{Rumor}}& \multicolumn{2}{c}{\textbf{Virality}}& \multicolumn{2}{c}{\textbf{Vul.}}\\ \cmidrule(lr){3-4}\cmidrule(lr){5-6}\cmidrule(lr){7-8}
&&Acc&MacF1& MSE& nDCG& MSE& nDCG\\\midrule
STL&-&0.806&0.796&0.209&0.990&0.122&0.972  \\\midrule
\multirow{3}*{MTL}&Basic&0.793&0.784&0.228&0.981&0.126&0.973\\
&CON& 0.820&0.833& 0.320&0.996&0.112&0.991\\
&Meta&\textBF{0.826}&\textBF{0.845}&\textBF{0.197}&\textBF{0.999}&\textBF{0.104}&\textBF{0.995}\\
\bottomrule
\end{tabular}
\label{tbl:neg_trans}
\end{table}

\subsubsection{Effect of Training Strategies}
While there is a correlation among tasks, training conflicts leading to negative transfer could occur if an appropriate training strategy is not selected, and our training method can mitigate this issue. Table~\ref{tbl:neg_trans} shows that when we train all three tasks with the basic multi-task training by linearly combining the losses of three tasks, the performance is lower than training them individually, suggesting a training conflict.
Gradnorm, used by concurrent training, obviously improves the basic training by adjusting the gradients to balance the learning rates of the different tasks. However, it only improves the performance over the single-task setting marginally. In contrast, the meta-learning strategy outperforms both the basic training and the Gradnorm approach on all three tasks, indicating it mitigates training conflict substantially. This superior performance might be attributed to the fact that in meta-learning, training conflicts have less chance to occur since only $\theta_b$ is shared by the different tasks.

When the training conflict is alleviated, that is, after finding a suitable training method, we find that the joint learning results are better than training them individually. Importantly, these results suggest that the proposed approach effectively captures the correlation between rumor virality and user vulnerability and utilizes this information to improve prediction performance. By identifying these correlations, the model can better predict how misinformation might spread and which users are more susceptible to it. Consequently, this allows for targeted interventions and strategies to prevent the spread of misinformation, thereby enhancing the overall effectiveness of combating fake news and rumors in online communities.

\subsubsection{User Community Visualization}
We visualize the three largest user communities using t-SNE~\citep{van2008visualizing} generated by the pooling layer of MT-META. As our model utilizes a soft assignment, we consider each user is assigned to the community with which they have the highest similarity. Figure~\ref{fig:visu} shows that our model can effectively pull users with similar embeddings together, 
which facilitates further utilization of information implied by the communities (e.g., the shared patterns and characteristics of users within each community) to help our tasks. 

\subsubsection{Case Study}
\begin{figure}
\centering
\subfigure[]{
\includegraphics[width=.4\linewidth]{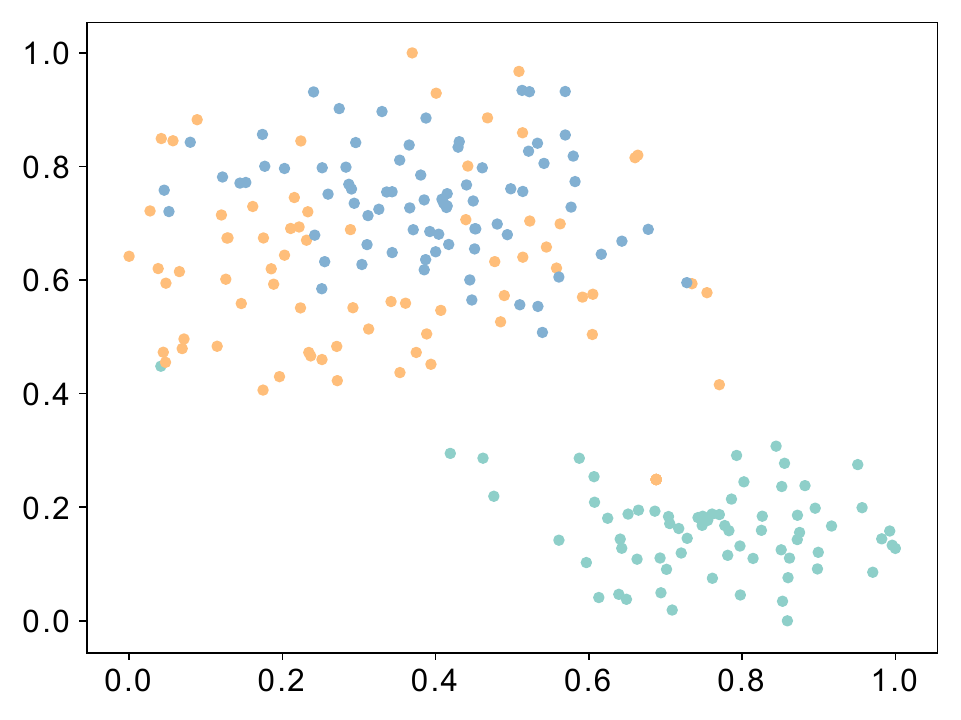}
\label{fig:visu}
}
\subfigure[]{
\includegraphics[width=.4\linewidth]{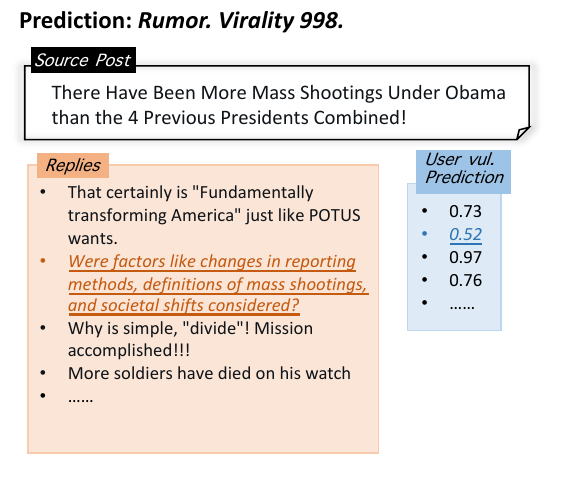} 
\label{fig:case}
}
\caption{(a) User community visualization, where each node is one user, and communities are colored differently.  (b) An example of a predicted viral rumor with user vulnerability scores by our model.}
\end{figure}

To better illustrate how our model leverages learning from the three tasks and utilizes the relationships between them to improve performance, we show an example in Figure~\ref{fig:case} using the graph of a viral rumor to interpret the prediction results.

We can see that our model predicts the event as a \emph{rumor} which appears consistent with its user vulnerability prediction, as most of the reposts are from vulnerable or moderate-vulnerable users. 
In addition, the event is unappealing to users that are moderately or not vulnerable (i.e., vulnerability score lower than 0.5).
The high user vulnerability scores highlight that people are easily swayed by the rumor and may readily accept and spread it without verification, which can explain why the rumor gets virally spread.
Compared to vulnerable users, the less vulnerable ones (e.g., with the post marked in orange color) seem to be more critical towards the claim as they notice the controversies of the source post instead of simply agreeing or repeating what others narrate. This observation conforms to the definition of user vulnerability: lower user vulnerability means the user is engaged more often in non-rumors (i.e., verified news) than rumors (i.e., false or unverified information), suggesting that this kind of user is more credible or rational.

\section{Implications}

Rumors are a constantly evolving phenomenon that requires collaborative efforts to mitigate their negative impact. For the technical contribution, we provide a new perspective, which combines three tasks of rumor detection, virality prediction, and user vulnerability scoring for infodemic surveillance. In this scenario, it proposes a mechanism that leverages the power of GNN to simultaneously learn these three tasks that were previously learned independently and captures the potential correlations among them. This not only improves the performance of individual tasks but also facilitates timely and effective information infodemic surveillance, that is, \textit{providing prompt and accurate responses to surveillance demands with limited information available}. Our framework also provides a new avenue for future research on infodemic surveillance.

For practical contribution, we provide a feasible strategy for infodemic surveillance in online social media platforms. 
The surveillance system first detects events and their spread on social media. The detected viral rumors are then sent to fact-checkers  for verification, and vulnerable users involved in rumors' spreading are also alerted to make them aware. In this way, firstly, vulnerable users can be precisely protected without disturbing the experience of other users as much as possible. Social media platforms are a crucial source of information and communication for many people. Disrupting the experience of other users could lead to confusion and mistrust, which could exacerbate the problem of infodemics.
Therefore, our approach is designed to be precise in identifying and protecting vulnerable users. This targeted approach enables us to provide protection to those who need it most while minimizing any negative impact on the broader user base.

Secondly, the burden of verification can be reduced, resulting in more effective information surveillance and early intervention. The sheer volume of information that is circulated on social media platforms makes it challenging to verify rumors and misinformation manually. Our unified prediction framework, which combines rumor detection, virality prediction, and user vulnerability scoring, enables a more comprehensive and accurate analysis of online information. This means that we can predict the most impactful rumors and vulnerable users simultaneously, which reduces the need for time-consuming manual verification, and has the potential to assist authorities in allocating resources more effectively.

Thirdly, one of the most significant implications of our unified prediction framework is its ability to detect rumors and predict their virality at an early stage, even when propagation information is limited. This is a crucial function of the model as it allows for timely intervention and prevents rumors from spreading further, potentially causing harm. Early detection of rumors and their potential impact enables authorities and organizations to respond quickly and effectively, thereby minimizing the risk of social unrest, political instability, and other adverse consequences.

\section{Conclusion and Future Work}
We propose a joint learning method for detecting rumors, and predicting their virality and user vulnerability in a unified multi-task framework based on graph neural networks. By leveraging the latent correlations of these tasks, our method can forecast rumors that potentially go viral and help find credulous users with a high propensity of spreading rumors, for timely and effective infodemic surveillance. 
The evaluation confirms that our method outperforms state-of-the-art baselines on all three tasks using two datasets with the ground truth of rumorous class, event virality, and user vulnerability constructed based on existing rumor detection corpora.

In the future, we plan to use a better ranking algorithm to replace the regression for further boosting nDCG. We will also develop approaches for embedding deeper user traits to better reflect users' internal states (maybe at a psychological level) towards rumors for more in-depth user vulnerability analysis. 

\section*{Declaration of competing interest}
The authors declare that they have no known competing interests that may influence the work reported in this paper.

\section*{Acknowledgement}
This research is supported by the
Singapore Ministry of Education (MOE) Academic Research Fund (AcRF) Tier-1 grant (Grant No. 19-C220-SMU-013). Any opinions, findings and conclusions or recommendations expressed in this material are those of the authors and do not reflect the views of funding agencies.
 
\appendix

\printcredits

\bibliographystyle{cas-model2-names}

\bibliography{cas-refs}


\end{document}